\documentclass{aa}

\usepackage{graphicx}
\usepackage{txfonts}

\newcommand{\Mpc}{$h^{-1}$\thinspace Mpc}

\newcommand{\etal}{{\rm et al.~}}

\def\apjl{ApJL}
\def\apjs{ApJS}

\def\aap{A\&A}

\begin{document}   

\title{Superclusters of galaxies in the 2dF redshift survey.  \\
III. The properties of galaxies in superclusters} 

\author{ M. Einasto\inst{1} \and J. Einasto\inst{1}\and E. Tago\inst{1} 
\and E. Saar\inst{1} \and L.J. Liivam\"agi\inst{1}
\and M. J\~oeveer\inst{1} \and  G. H\"utsi\inst{1} \and
P. Hein\"am\"aki\inst{3} 
\and V. M\"uller\inst{4} 
 \and D. Tucker\inst{5} 
}

\institute{Tartu Observatory, EE-61602 T\~oravere, Estonia
\and 
Tuorla Observatory, V\"ais\"al\"antie 20, Piikki\"o, Finland 
\and
Astrophysical Institute Potsdam, An der Sternwarte 16,
D-14482 Potsdam, Germany
\and
 Fermi National Accelerator Laboratory, MS 127, PO Box 500, Batavia,
IL 60510, USA
}

\date{ Received 2006; accepted} 

\authorrunning{M. Einasto et al.}

\titlerunning{2dFGRS supercluster galaxies}

\offprints{M. Einasto }

\abstract
 {}
{We use catalogues of superclusters of galaxies from the 2dF 
Galaxy Redshift Survey to study the properties of galaxies in superclusters. }
{We compare the properties 
of galaxies in high and low density regions of rich superclusters, in poor 
superclusters and in the field, as well as 
in groups, and of isolated galaxies in superclusters of various richness.}
{We 
show that in rich superclusters
the values of the luminosity density  smoothed on 
a scale of 8 \Mpc\   are higher than 
in poor superclusters: the median density in rich superclusters is 
$\delta \approx 7.5$, in poor superclusters $\delta \approx 6.0$. Rich 
superclusters contain high density cores with densities $\delta > 10$ while in 
poor superclusters such high density cores are absent. The properties of 
galaxies in rich and poor superclusters and in the field are different: the 
fraction of early type,  passive galaxies in rich superclusters is slightly 
larger than in poor superclusters, and is the smallest among the field galaxies. 
Most importantly, in high density cores of rich superclusters ($\delta > 10$) 
there is an excess of early type, passive galaxies in 
groups and clusters, as well as among those
which do not belong to groups or clusters. 
The main galaxies of superclusters have a 
rather limited range of absolute magnitudes.  The main galaxies of rich 
superclusters have larger  luminosities  than those of poor 
superclusters and of groups in the field (the median values  are 
correspondingly $M_{bj} = -21.02$, $M_{bj} = -20.9$  and $M_{bj} = -19.7$
for rich and poor superclusters and  groups in the field).}
{Our results show that both the local (group/cluster)
environments and global (supercluster) environments influence
galaxy morphologies and their star formation activity.

}

\maketitle

\section{Introduction}

It is presently well established that galaxies belong to various systems from
groups and clusters to superclusters, forming the supercluster-void network.
Early studies of superclusters of galaxies were reviewed by Oort
(\cite{oort83}). These studies were based on observational data about
galaxies, as well as on data about nearby groups and clusters of galaxies.
Classical cluster catalogues were constructed by Abell (\cite{abell}) and
Abell \etal (\cite{aco}) by visual inspection of Palomar plates. The first
relatively deep all-sky catalogues of superclusters of galaxies were complied
by Zucca \etal (\cite{z93}) and Einasto \etal (\cite{e1994}, \cite{e1997}, 
\cite{e2001}) using data about Abell clusters.

The modern era of the study of various systems of galaxies began when new deep
redshift surveys of galaxies were published. These surveys cover large
regions of sky and allow to investigate the distribution of galaxies up to
fairly large distance from us. These surveys formed the basis for new
catalogues of groups, clusters and superclusters of galaxies.  The first of
such catalogues was the Las Campanas catalogue of groups by Tucker \etal
(\cite{Tucker00}).  The Las Campanas Galaxy Redshift Survey and the Sloan
Digital Sky Survey were also used to compile catalogues of groups, clusters
and superclusters by Einasto et al. (\cite{e03a}, \cite{e03b}, hereafter E03a
and E03b) and Basilakos (\cite{bas03}).  Group and supercluster catalogues
based on the 2 degree Field Galaxy Redshift Survey (2dFGRS) were published by
Eke \etal (\cite{eke04}), Yang \etal (\cite{yang04}) and Tago et al.
(\cite{tago06}, hereafter T06), and by Erdogdu et al. (\cite{erd04}) and
Porter and Raychaudhury (\cite{pr05}).

The pioneering studies of the properties of galaxies in clusters by Davis \&
Geller (\cite{dg76}) and Dressler (\cite {d80}) showed that there exists a
correlation between the spatial density of galaxies and their morphology --
early type galaxies are located preferentially in the central regions of
clusters, where the local densities are high, while late type galaxies are
located mostly in outer regions of clusters, having lower local densities 
around them.
Einasto (\cite{e91}) showed that clustering of galaxies depends on both
their luminosity and morphology.  Already early studies of the morphological
segregation of galaxies at supercluster scales demonstrated that this
segregation extends to scales of 10--15 \Mpc\ 
(Giovanelli, Haynes and Chincarini \cite{gio86}, Einasto and Einasto
\cite{e87} and Mo et al.  \cite{mo92}).

The data about galaxies in the Las Campanas Galaxy Redshift Survey, the Sloan Digital
Sky Survey and the 2 degree Field Galaxy Redshift Survey (2dFGRS) enable us to study
the properties and the spatial distribution of galaxies in detail. Numerous papers
have demonstrated segregation of galaxies by their spectral
type, luminosity and colour index (Norberg et al. \cite{n01}, \cite{n02},
Zehavi et al. \cite{zehavi02}, Goto et al. \cite{go03}, Hogg et al.
\cite{hogg03}, \cite{hogg04}, Balogh et al. \cite{balogh04}, De Propris et al.
\cite{depr03}, Madgwick et al. \cite{ma03b}, Croton et al. \cite{cr05} and
Blanton et al. \cite{blant04}, \cite{blant06} among others).  Blanton et al.
(\cite{blant06}) come to the conclusion that the blue galaxy fraction and the
recent star formation history in general, depend mostly on the local
environment of galaxies.  In this study the local environment was defined as
the spatial density on the 0.5--1~\Mpc\ scale, and the global density on the
5--10~\Mpc\ scale.  Croton et al. (\cite{cr05}) showed that galaxy populations
depend also on the large scale environment.
Balogh et al. (\cite{balogh04}) compared the populations of
star-forming and quiescent galaxies in groups from the 2dFGRS and SDSS surveys
and in small (1.1~\Mpc) and large scales (5.5~\Mpc)
 and showed that the relative numbers of these 
galaxies depend both on the local and global environments. Even low density environments 
contain a large fraction of non-star-forming galaxies.
 
On the basis of the 2dFGRS we recently compiled a new catalogues of superclusters
of the 2dF galaxies (Einasto et al. \cite{e06a}, hereafter Paper I).  
This catalogue  is available electronically at the web-site
\texttt{http://www.aai.ee/$\sim$maret/2dfscl.html}.  In (Einasto et al.
\cite{e06b}, hereafter Paper II) we studied various properties of these
superclusters: their multiplicity, geometry, luminosity functions and other
properties. We also compared the properties of real superclusters with
simulated superclusters from the Millennium Simulations (Springel et al. 
\cite{springel05}), and from the semianalytical mock catalog by Croton et al. 
(\cite{cr06}).

In the present paper we continue our study of the properties of superclusters. 
We study the properties of galaxies: their  luminosities, spectral types and 
colors in rich and poor superclusters  and for comparison also in the field. 
These data enable to analyse populations of galaxies of different luminosity, 
morphology and star formation rate in various environments: in rich and poor 
superclusters, as well as in groups located in superclusters and in the field. 
The use of a large catalogue of superclusters enables us for the first time to 
study the properties of galaxies in a large number of superclusters of various 
richness.

The paper is composed as follows.  In the next Section we shall describe the
supercluster data. Then we study  the properties of galaxies in
superclusters, the density distribution in superclusters
of various richness, and the properties of 
galaxies in groups located in regions of different large scale density
in superclusters. Then we compare
the luminosities of main galaxies in superclusters and in groups 
located in the field. In
the last Sections we discuss the results and list our conclusions.

\section{Data}

We have used the 2dFGRS final release (Colless \etal \cite{col01},
\cite{col03}) that contains 245,591 galaxies. This survey has allowed the
2dFGRS Team and many others to estimate fundamental cosmological parameters
and to study intrinsic properties of galaxies in various cosmological
environments;  see Lahav (\cite{lahav04} and \cite{lahav05}
) for recent reviews.

\begin{figure}[ht]
\centering
\resizebox{0.48\textwidth}{!}{\includegraphics*{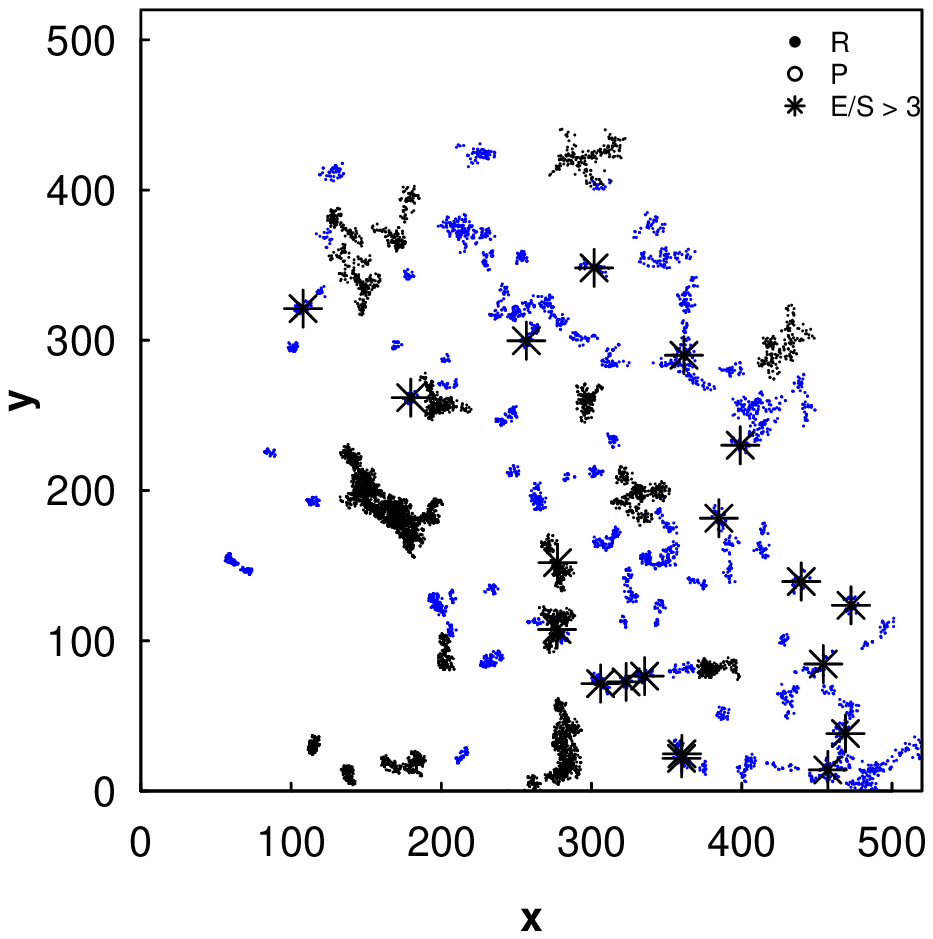}}
\hspace{2mm} 
\raisebox{2pt}{\resizebox{0.48\textwidth}{!}{\includegraphics*{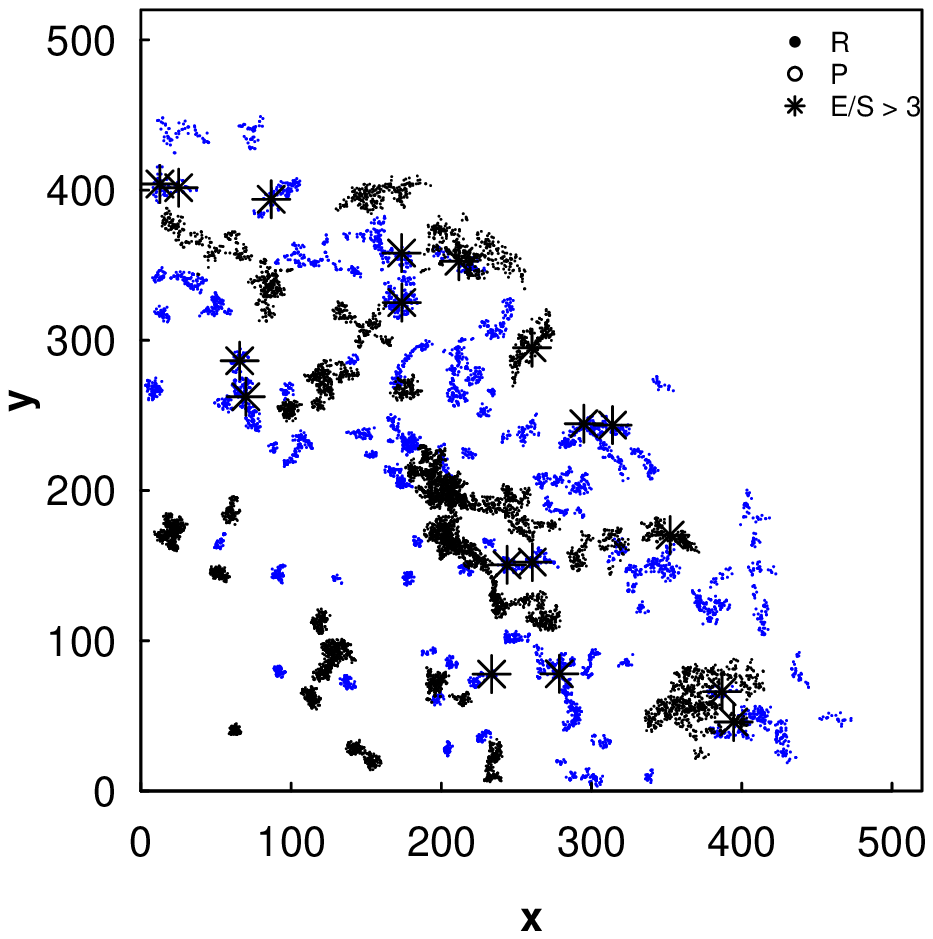}}}
\caption{Galaxies in superclusters with at least 20 member galaxies. 
Upper panel: Northern sky, lower panel: Southern sky. 
Darker dots represent galaxies in rich superclusters
with at least 200 member galaxies, 
lighter dots - galaxies in poor superclusters. Stars indicate  
poor superclusters, which have the ratio $E/S \geq 3$ (see Section 3.1).}
\label{fig:1}
\end{figure}

\begin{figure}[ht]
\centering
\resizebox{0.48\textwidth}{!}{\includegraphics*{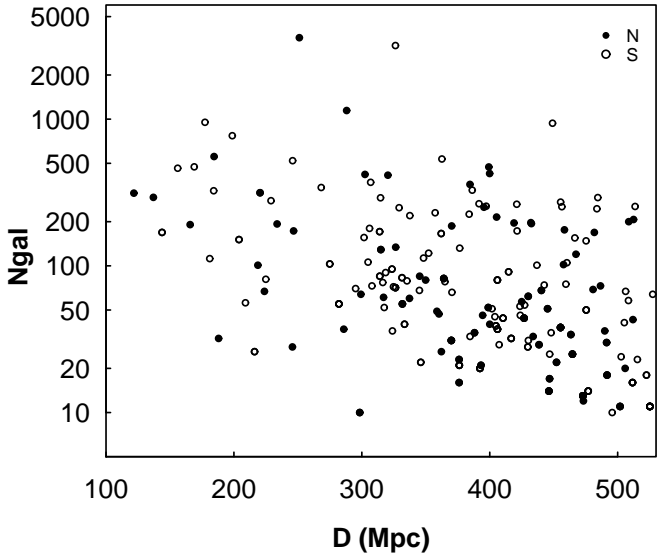}}
\caption{The distribution of the number of galaxies in
  superclusters at various distance. N - Northern sky, S - Southern sky. 
}
\label{fig:2}
\end{figure}

We used the data about galaxies and groups of galaxies to compile a 
catalogue of superclusters of galaxies from the 2dF survey (Paper I). The 
2dF sample becomes very diluted at large distances, thus we restrict our 
sample by a redshift limit $z=0.2$; we apply a lower limit $z \geq 
0.009$ to avoid confusion with unclassified objects and stars. We 
determined superclusters using a luminosity (or mass) density field 
smoothed with an Epanechnikov kernel of radius 8~\Mpc\
($h$ is the present-day dimensionless Hubble
constant in units of 100 km s$^{-1}$ Mpc$^{-1}$).
When  calculating luminosity 
densities we used weights for galaxies which correct the densities to
account for galaxies, 
too faint to fall into the observational window of 
absolute magnitudes at the distance of the galaxy. We defined 
superclusters as connected non-percolating systems with densities above 
a certain threshold density.  We used a threshold density 4.6 in units of 
the mean luminosity density. 

Due to the selection of galaxies within a fixed apparent magnitude limits the
observational window in absolute magnitudes shifts toward higher luminosities
when the distance of galaxies increases.
We analysed the selection effects in our
supercluster catalogue in detail in Papers I and II.  This analysis showed, in
particular, that selections due to the use of a flux-limited sample of galaxies
have been taken into account properly when estimating  total luminosities
of superclusters.
For details we refer to Papers I and II.

We use
in the present analysis only systems with at least 10 member galaxies from
main supercluster catalogue of Paper I, which includes all systems up to a
comoving distance $D = 520$ \Mpc.  In total we use the data about 124 superclusters
in the Northern sky and about 168 superclusters in the Southern sky,
leaving out the poorest systems in the supercluster catalogue.  

Another selection effect in flux-limited
samples is the decrease of the number of galaxies in superclusters with
increasing distance (Fig.~\ref{fig:2}).  This selection effect affects 
poor superclusters
at distances larger than $D > 300$\Mpc\ more strongly than 
superclusters at smaller distances.  Therefore
we divide our superclusters into nearby and distant subsamples and 
select volume-limited samples as follows:  nearby samples -- $M_{bj}
\leq -18.4$  (in the $b_j$ filter used in the 2dFGRS) and $D \leq
300$\Mpc (denoted with $N$), and distant samples having
$M_{bj} \leq -19.7$ and $D > 300$\Mpc (denoted with $D$).

We additionally divide superclusters by their richness: rich 
superclusters with at least 200 member galaxies (we denote this sample 
as R),  and poor superclusters with less than 200 member galaxies 
(P). We also use the data about field galaxies, i.e. galaxies which do not 
belong to superclusters (approximately 2/3 of all galaxies), as a 
comparison sample (FG).  Our analysis showed  that for the purposes of 
the present paper this division of superclusters into rich and poor 
systems is better than the division according to the number of DF-clusters 
used in Papers I and II (Section 3.2).

To study various  properties of galaxies in superclusters we use the data about 
luminosities, spectral types and colors of galaxies as given in the 2dF redshift 
survey.   We divide galaxies into populations of bright/faint galaxies, 
early/late type galaxies, non-star-forming and star-forming galaxies and 
passive/actively star formating galaxies by their luminosity, spectral parameter 
$\eta$ and by the colour index (Madgwick et al. \cite{ma02} and \cite{ma03a}, 
Wild et al. \cite{wild04}). 

In order to divide galaxies into populations of bright and faint galaxies we 
wanted to use an absolute magnitude limit close to the break luminosity $M^*$ in 
the Schechter luminosity function. According to the calculations of the 
luminosity function the value of $M^\star$ varies for different galaxy populations 
((Madgwick et al. \cite{ma03a}, de Propris et al. \cite{depr03}, Croton et al. 
\cite{cr05}), having values from $-19.0$ to $-20.9$. Therefore we used a 
bright/faint galaxy limit $M_{bj} = -20.0$  as a compromise between different 
values.

The spectral parameter of galaxies, $\eta$, correlates 
with the morphological type of galaxies (e.q. 
Madgwick et al. \cite{ma02}, de Propris et al. \cite{depr03});
E/S0 galaxies (morphological T-type $T < 0$, Kennicutt \cite{ken92})
have $\eta \leq -1.4$. Thus we divided galaxies into populations
of early and late type galaxies using this limit of the spectral
parameter $\eta$. Moreover, the spectral parameter $\eta$ is correlated with
the equivalent width of the $H_{\alpha}$ emission line, thus being 
an indicator of the star formation rate in galaxies 
(Madgwick et al. \cite{ma02} and \cite{ma03a}). We used 
the value $\eta < 0.0$ to define the population of
quiescent galaxies and $\eta \geq 0.0$ for star-forming galaxies.

We also used information about colours of galaxies (the rest-frame colour 
index, $col = (B - R)_0$, Wild et al. \cite{wild04}) to divide galaxies into 
populations of passive galaxies and actively star forming galaxies. For passive 
(red) galaxies $col \geq 1.07$. We used this limit to separate the populations of 
passive and actively star forming galaxies.

In addition to these spectral parameters of galaxies we use the
data about groups of galaxies (T06) to find 
the fraction of galaxies in groups of various richness and to study
the properties of galaxies in groups in rich and poor superclusters.

\section{Properties of galaxies in superclusters}

{\scriptsize
\begin{table}[ht]
\caption{The galaxy content in superclusters. }
\begin{tabular}{lrrrrrr} 
\hline 
\\ 
ID     &    R$_N$   & R$_D$ &  P$_N$ &  P$_D$    & FG$_N$   &  FG$_D$       \\
1      &    2       & 3     &  4     &  5        & 6       &  7            \\
\hline 
\\ 
$N_{gal}$ &&&& \\
\emph{All}  &  7461   & 5747  &   1652  &  7031   & 36949  &  32162  \\
$B/F$  &  0.17   & 1.86  &   0.18  &  2.21   &  0.14  &   1.55  \\
\\
$E/S$ &&&& \\
\emph{All}    &  1.23   & 1.78  &  1.21  &  1.77  &  0.66  &  1.02    \\
$B$      &  2.66   & 2.16  &  2.36  &  2.15  &  1.60  &  1.21     \\
$F$      &  1.09   & 1.27  &  1.08  &  1.18  &  0.58  &  0.78     \\
\\
 
$q/${\it SF} &&&& \\

\emph{All}    &  2.42   & 3.63  &  2.29  &  3.61  &  1.41  &  2.24    \\
$B$      &  7.31   & 4.43  &  6.78  &  4.47  &  4.34  &  2.64     \\
$F$      &  2.11   & 2.63  &  1.98  &  2.41  &  1.24  &  1.76     \\
\\
 
$P/A$ &&&& \\

\emph{All}    &  1.99   & 3.27 &  1.98  &   3.72  &  1.03  &  2.24  \\
$B$      &  3.11   & 3.69 &  2.52  &   4.42  &  1.59  &  2.48  \\
$F$      &  1.84   & 2.65 &  1.89  &   2.67  &  0.97  &  1.93  \\
\\
 
$F_{gr}$ &&&& \\

$Gr_{10}$&  0.34   &  0.09 &  0.30  &  0.05  &  0.05  &  0.003       \\
$Gr_2$   &  0.41   &  0.47 &  0.47  &  0.48  &  0.43  &  0.29       \\
$I$      &  0.25   &  0.45 &  0.23  &  0.47  &  0.52  &  0.71       \\
\\
\hline 
\label{tab:1}
\end{tabular}
{

The Columns in the Table are as follows:

\noindent  1: Sample identification (all -- all galaxies, $B$ -- bright galaxies, 
$F$ -- faint galaxies. $B/F$ -- the ratio of the numbers of bright 
($M_{bj}\leq -20.0$) and faint ($M_{bj} > -20.0$) galaxies, 
$E/S$ -- the ratio  of the number of early and late type galaxies, $q/SF$ -- 
the ratio of the numbers
of quiescent and actively star forming  galaxies (according to the spectral 
parameter $\eta$), $P/A$ --  the ratio of the number of passive and actively 
star forming galaxies (according to the colour index $col$).
$F_{gr}$ -- the fraction of galaxies in groups; $Gr_{10}$ -- rich groups
with at least ten member galaxies, $Gr_2$ -- poor groups with less 
than ten galaxies, and $I$ -- isolated galaxies, i.e. those galaxies
which do not belong to groups.

\noindent  2--7: Supercluster populations: 
R$_N$ and R$_D$ -- nearby and distant rich superclusters, 
P$_N$ and P$_D$ -- nearby and distant poor superclusters, 
FG$_N$ and FG$_D$ -- nearby and distant field galaxies.
}

\end{table}
}

{\scriptsize
\begin{table}[ht]
\caption{The galaxy content in superclusters. 
The Kolmogorov-Smirnov test results.}
\begin{tabular}{lrrr} 
\hline 
\\ 
ID1       &    ID2   &    $D$     &  $P$   \\
1       &    2   &    3     &  4   \\
\hline 
\\ 
$Bmag$     &         &         &     \\
\hline 
R$_N$      & P$_N$   &   0.040   &     0.02635       \\
R$_N$      & FG$_N$  &   0.036   &     1.076e-07     \\
P$_N$      & FG$_N$  &   0.069   &     4.117e-07     \\
   \\                       
R$_D$      & P$_D$   &   0.041   &    3.594e-05       \\
R$_D$      & FG$_D$  &   0.074   &    $<$ 2.2e-16         \\
P$_D$      & FG$_D$  &   0.100   &    $<$ 2.2e-16         \\
    \\                      
\hline 
\\ 
$col$     &         &         &     \\
\hline 
R$_N$      & P$_N$   &   0.061   &     8.495e-05     \\
R$_N$      & FG$_N$  &   0.162   &     $<$ 2.2e-16       \\
P$_N$      & FG$_N$  &   0.173   &     $<$ 2.2e-16       \\
   \\                       
R$_D$      & P$_D$   &   0.049   &    2.839e-07     \\
R$_D$      & FG$_D$  &   0.118   &    $<$ 2.2e-16           \\
P$_D$      & FG$_D$  &   0.135   &    $<$ 2.2e-16           \\
    \\                      
\hline 
\\ 
$\eta$     &         &         &     \\
\hline 
R$_N$      & P$_N$   &   0.027   &     0.2564      \\
R$_N$      & FG$_N$  &   0.158   &     $<$ 2.2e-16     \\
P$_N$      & FG$_N$  &   0.156   &     $<$ 2.2e-16     \\
   \\                       
R$_D$      & P$_D$   &   0.041  &   6.066e-05     \\
R$_D$      & FG$_D$  &   0.136  &    $<$2.2e-16           \\
P$_D$      & FG$_D$  &   0.135  &    $<$2.2e-16           \\
    \\                      
\\ 
\hline 
\label{tab:2}
\end{tabular}
{

The Columns in the Table are as follows:

\noindent  1--2: Sample ID (see Table~\ref{tab:1}). 

\noindent  3--4: the Kolmogorov-Smirnov test results:
the maximum difference and the probability that the 
distributions of population parameters are taken
from the same parent distribution.
}

\end{table}
}

\subsection{Luminosities, types and colours 
of galaxies in superclusters of different richness}

\begin{figure*}[ht]
\centering
\resizebox{0.45\textwidth}{!}{\includegraphics*{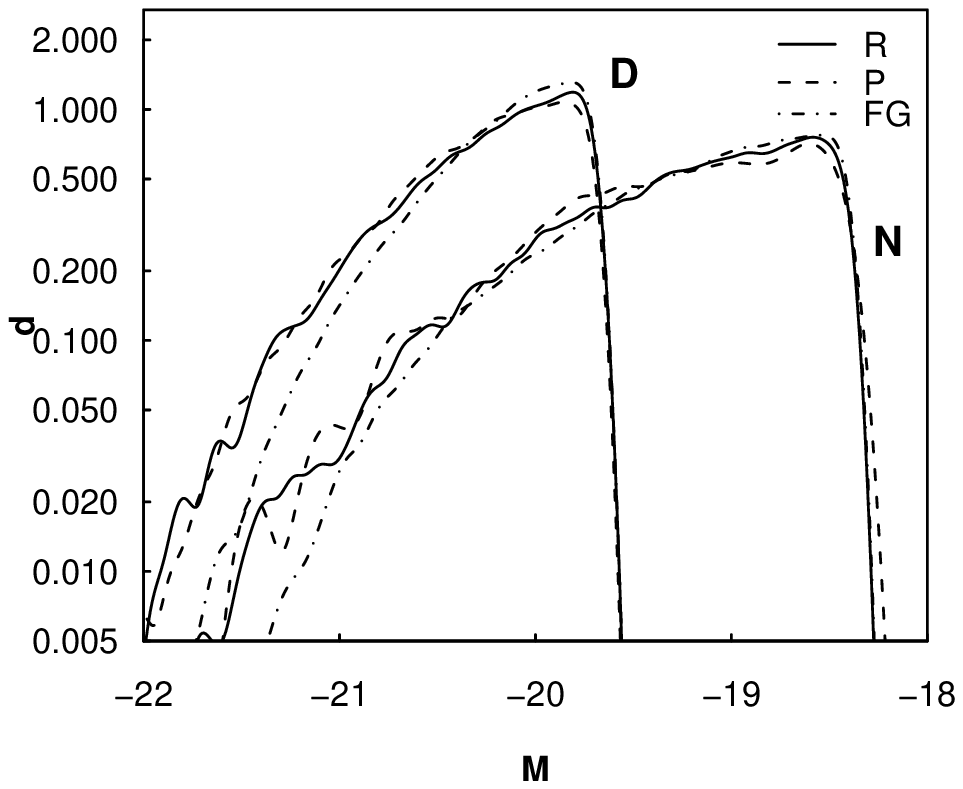}}
\resizebox{0.45\textwidth}{!}{\includegraphics*{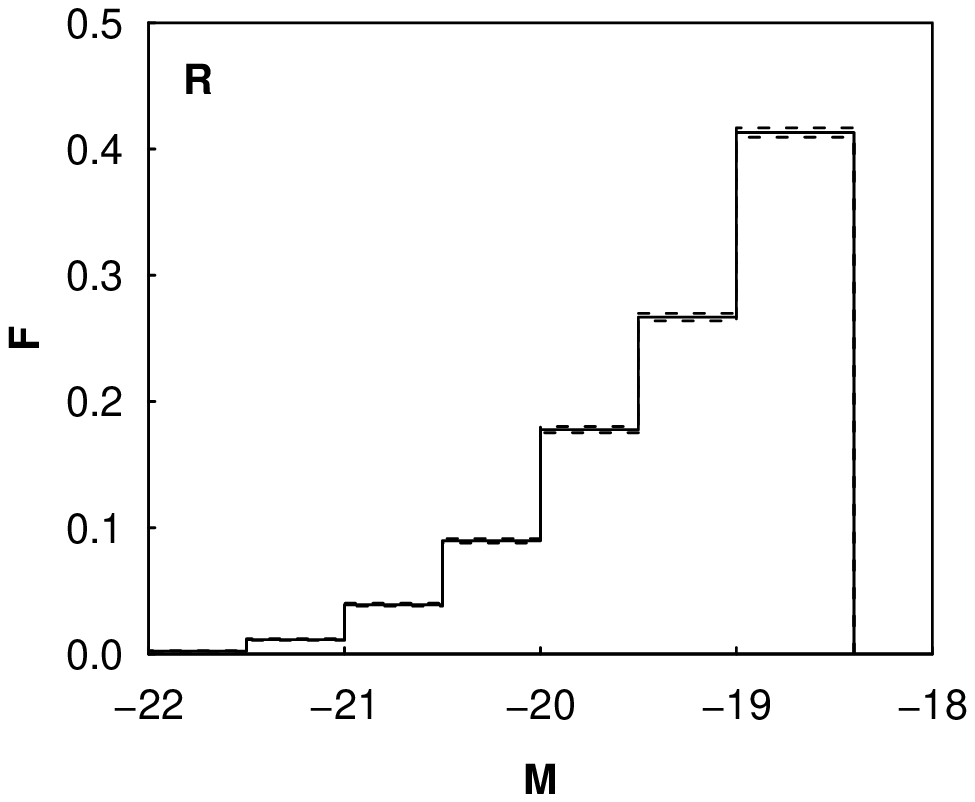}}
\caption{Left panel: 
the differential luminosity functions for galaxies of the 2dFGRS  
in rich (R) and poor (P) superclusters and in field (FG)
for nearby (N) and distant (D) samples. 
Right panel: differential luminosity function 
$F=dN/dM$, where $M$ is the absolute magnitude of a galaxy,
for galaxies in rich (R) superclusters; here 
solid line show luminosity function and dashed lines indicate 
Poisson errors.}
\label{fig:3}
\end{figure*}

\begin{figure*}[ht]
\centering
\resizebox{0.28\textwidth}{!}{\includegraphics*{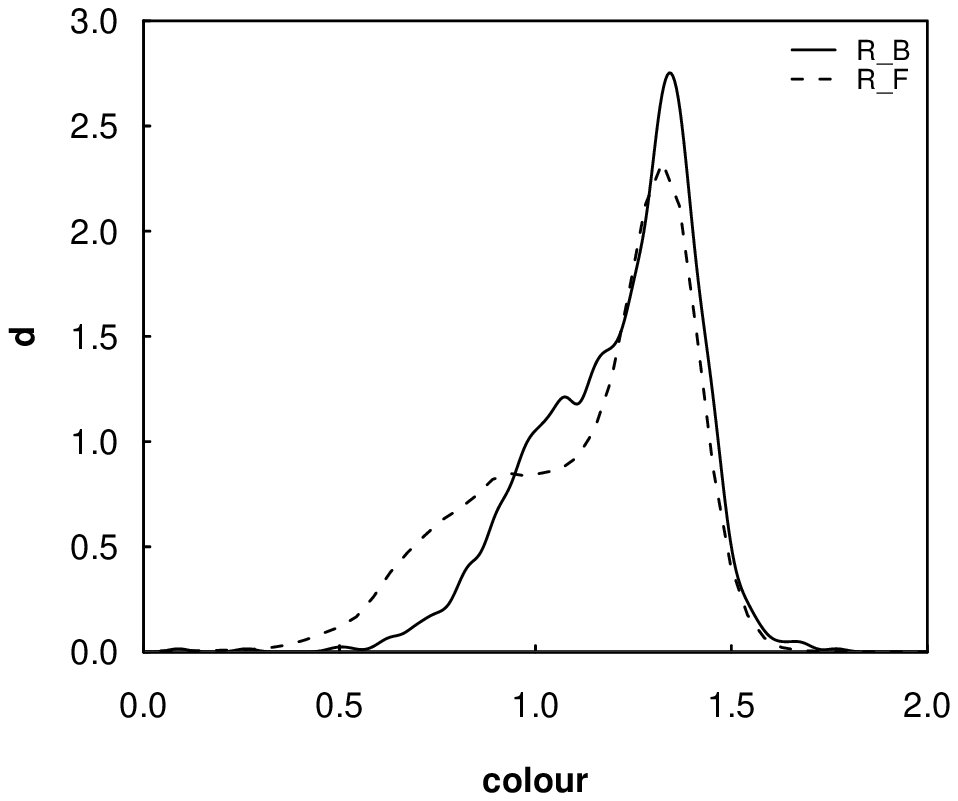}}
\resizebox{0.28\textwidth}{!}{\includegraphics*{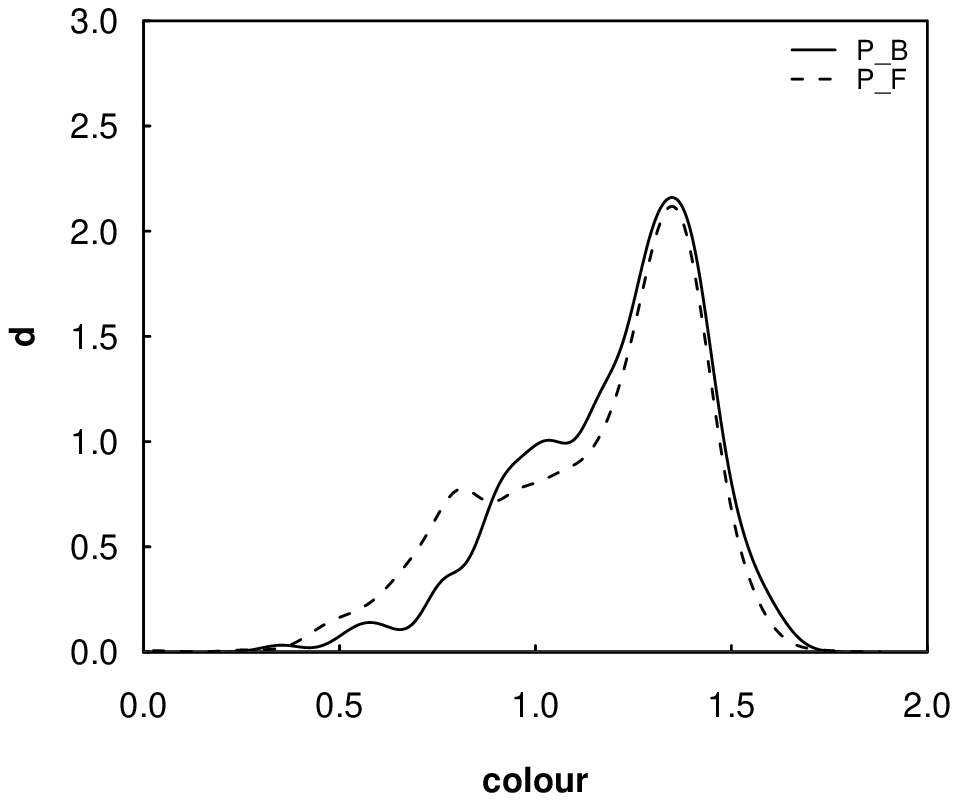}}
\resizebox{0.28\textwidth}{!}{\includegraphics*{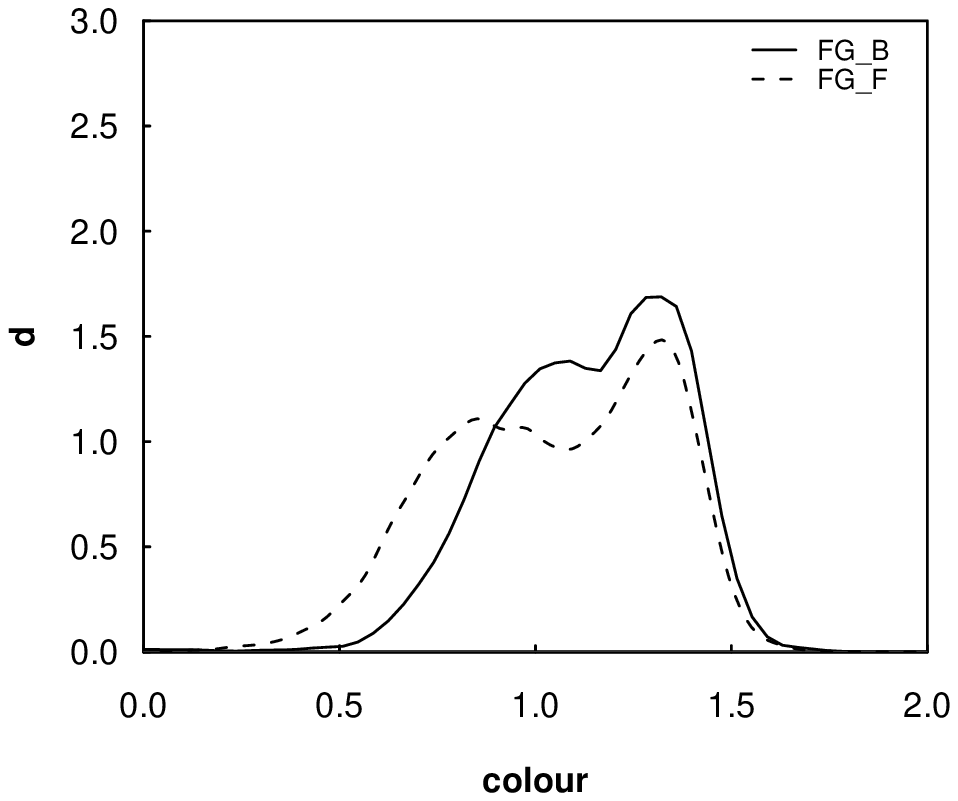}}
\hspace{2mm} \\
\raisebox{2pt}
{\resizebox{0.28\textwidth}{!}{\includegraphics*{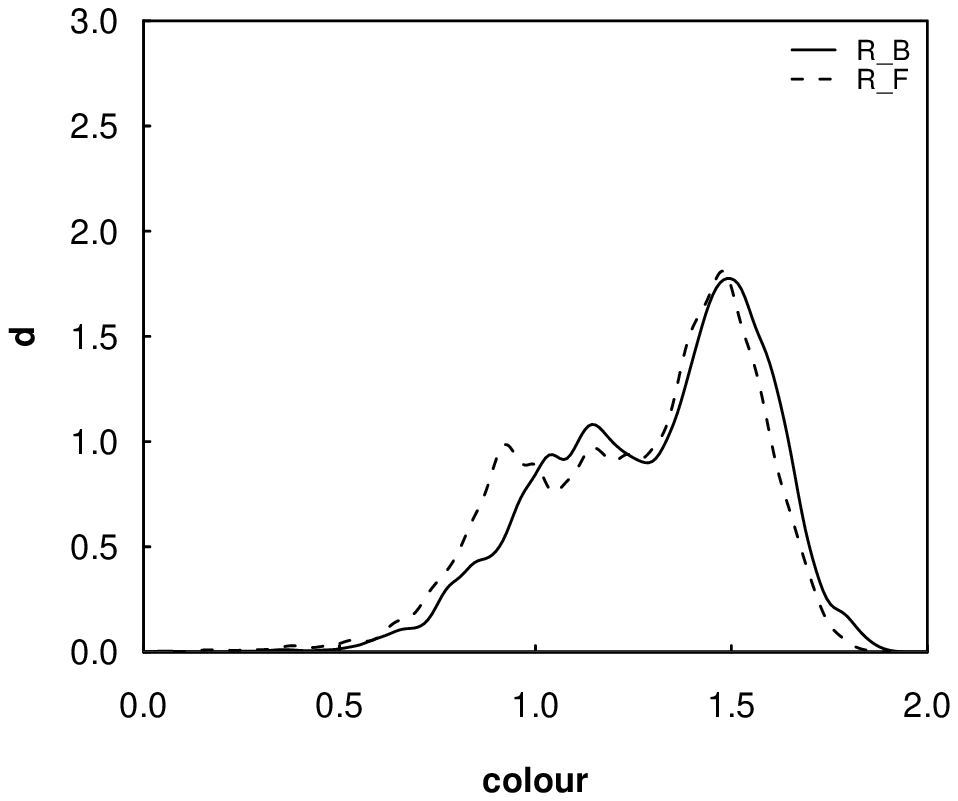}}}
{\resizebox{0.28\textwidth}{!}{\includegraphics*{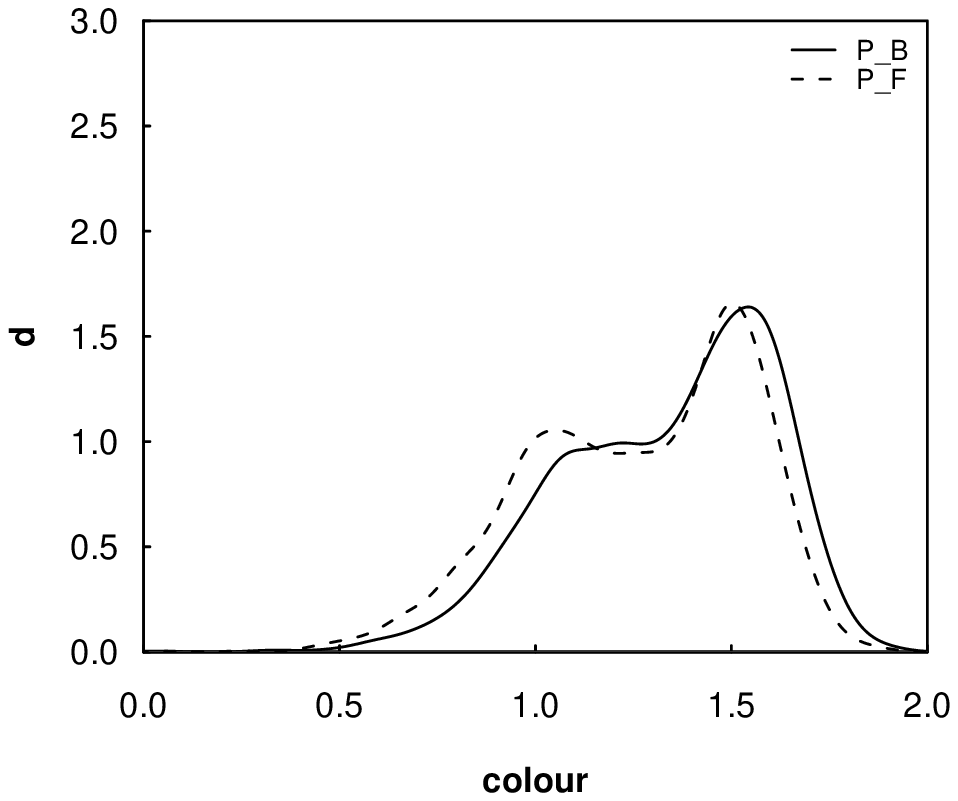}}}
{\resizebox{0.28\textwidth}{!}{\includegraphics*{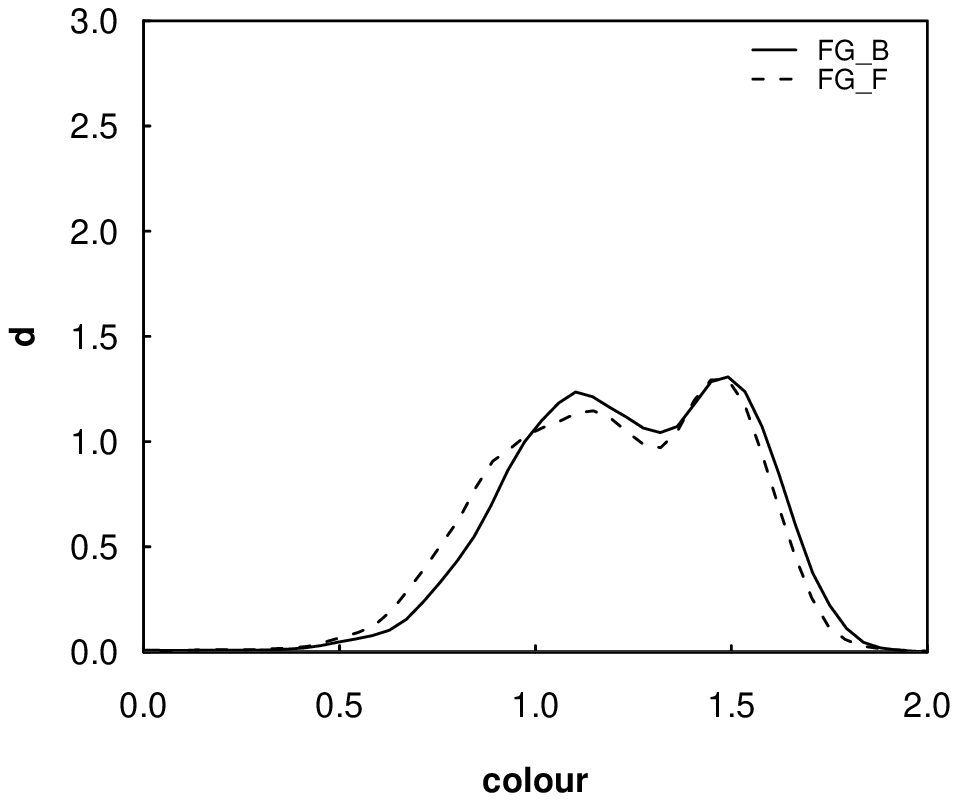}}}
\caption{The distribution of the colour index $col$ for galaxies
in rich (R, left panels) and poor (P, middle panels) superclusters 
and in the field (FG, right panels), volume-limited
samples, galaxies divided by luminosity: bright galaxies with 
$M_{bj} \leq -20.0$ and faint galaxies having $M_{bj} > -20.0$. 
Upper panels: $D \leq$ 300
\Mpc, lower panels: $D \ge$ 300 \Mpc. }
\label{fig:4}
\end{figure*}

Table~\ref{tab:1} shows the galaxy content of  rich and poor superclusters and 
field galaxies.  Here we divide galaxies also by their luminosity: bright 
galaxies with $M_{bj} \leq -20.0$ and faint galaxies having $M_{bj} > -20.0$. 
In Table~\ref{tab:2} we give the statistical significance that the distributions 
of luminosities, spectral parameter $\eta$ and colour index $col$ are drawn from 
the same parent sample according to the Kolmogorov-Smirnov test.  

We plot in Figure~\ref{fig:3} the differential luminosity 
functions for galaxies   
in rich and poor superclusters and in the field (right panel), as well as
the distribution of luminosities of
galaxies  in rich and poor superclusters and in the field (left panel). As we see,
the Poisson errors in the probability
density histograms are very small, due to a large number of galaxies
in our sample. Thus we do not show these errors, in general, to avoid
overcrowding of the figures.
Figure~\ref{fig:4} shows the distributions of the colour indices $col$
of galaxies. We calculated these distributions 
using the probability density function in
R (a language for data analysis and   graphics
(Ihaka \& Gentleman \cite{ig96}).

Fig. ~\ref{fig:3} shows that there is an excess of faint galaxies 
among field galaxies in comparison with galaxies 
in superclusters.  The ratio of the numbers of bright and faint galaxies in the 
field is smaller than this ratio for galaxies in superclusters 
(Table~\ref{tab:1}). The KS test shows (Table~\ref{tab:2}) that the differences
between the luminosity distributions of galaxies in rich and poor superclusters
are statistically significant at least at 97\% confidence level, 
and the differences 
between luminosity distributions of galaxies in superclusters and in the field --
even at a much higher level (Table~\ref{tab:1}).

The ratio of the numbers of early and late type galaxies, $E/S$
(characterized  by their spectral parameter $\eta$) in rich and poor superclusters
and in the field (Table~\ref{tab:1})
shows differences between these populations: this ratio is slightly larger
for galaxies in rich superclusters than 
for galaxies in poor superclusters. 
Note that this ratio is also larger for bright galaxies in rich
superclusters than for bright galaxies in poor
superclusters -- in rich superclusters the fraction of early
type galaxies among bright galaxies is larger than in poor superclusters.
In the field the fraction 
of early type galaxies is smaller than in superclusters.

The ratio of the numbers of quiescent and actively star forming galaxies,
$q/$\emph{SF}
(as defined by their spectral parameter $\eta$) in rich and poor superclusters
and in the field (Table~\ref{tab:1}) shows even larger differences between bright
and faint galaxies both in superclusters and in the field;
for bright galaxies this fraction is about three times higher
than for faint galaxies. In distant superclusters, where we use
a higher magnitude limit to define a volume-limited sample, the
difference between this fraction for bright and faint galaxies is smaller.

We note another difference between the galaxy populations of rich
superclusters and in the field -- in rich superclusters there is an excess of
early type (and passive) galaxies among faint galaxies while in the
field late type (and star forming) galaxies dominate among faint galaxies.

The ratio of passive and active galaxies $P/A$ according to their
colour index $col$ 
is given in Table~\ref{tab:1} for galaxies in rich and poor superclusters and 
in the field. In Figure~\ref{fig:4}
we show the distributions of the colour index  for bright and faint galaxies in these systems. This Figure shows a continous 
change of the distributions of colours of galaxies from rich to poor 
superclusters and to field galaxies; differences 
are larger in the case of nearby samples. 
The number of red galaxies in rich superclusters is larger than in poor superclusters
among both bright and faint galaxies.
In the case of the distant sample the colour of galaxies is redder.
We shall analyse this effect in more detail in another paper 
(Einasto et al., in preparation).
Thus in rich 
superclusters there is an excess of quiescent galaxies even among faint galaxies, 
while in the field the fraction of actively star forming galaxies is relatively 
large. The fraction of actively star forming galaxies in the field is larger 
than the fraction of these galaxies in superclusters.

The KS test shows that the differences between the distributions of colour
indices of galaxies in both nearby and distant rich and poor superclusters, as
well as the differences between colour indices of galaxies in superclusters
and in the field are statistically significant at least at 99\% confidence level.

The differences between the distributions of the spectral parameters
and colour indices of bright and faint galaxies are expected due 
to the morphology-luminosity-density relation. However, our results 
show that galaxy populations in rich and poor superclusters are somewhat
different; in rich  supercluster there are relatively more early type, passive,
red galaxies than in poor superclusters.

\subsection{Properties of galaxies in individual superclusters}
\begin{figure*}[ht]
\centering
\resizebox{0.28\textwidth}{!}{\includegraphics*{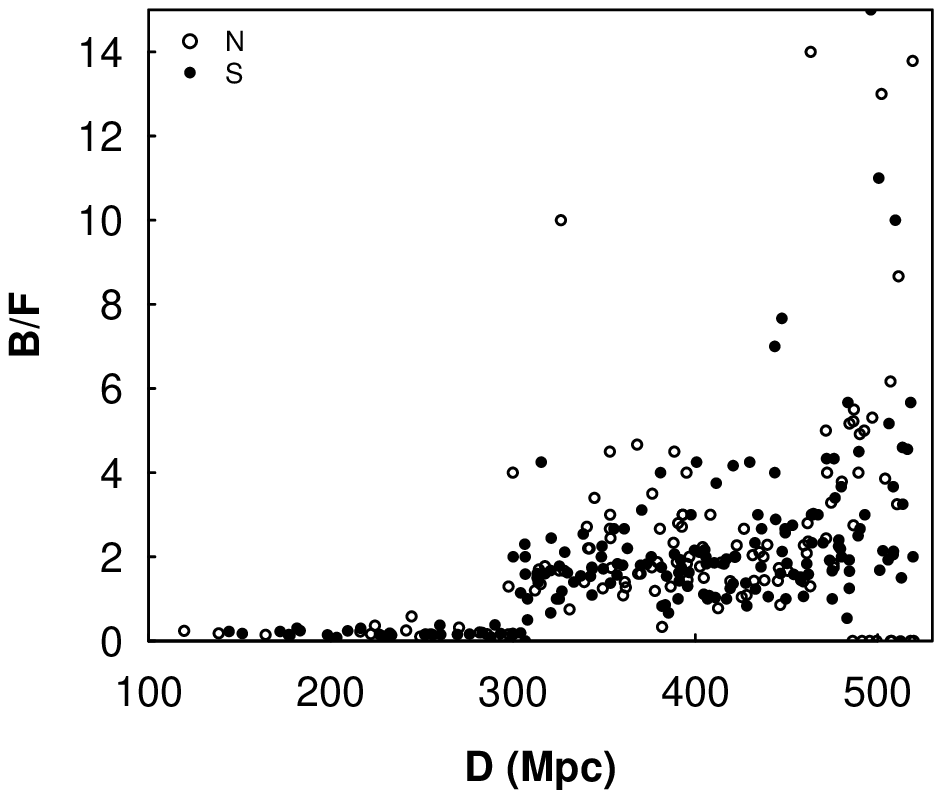}}
\resizebox{0.28\textwidth}{!}{\includegraphics*{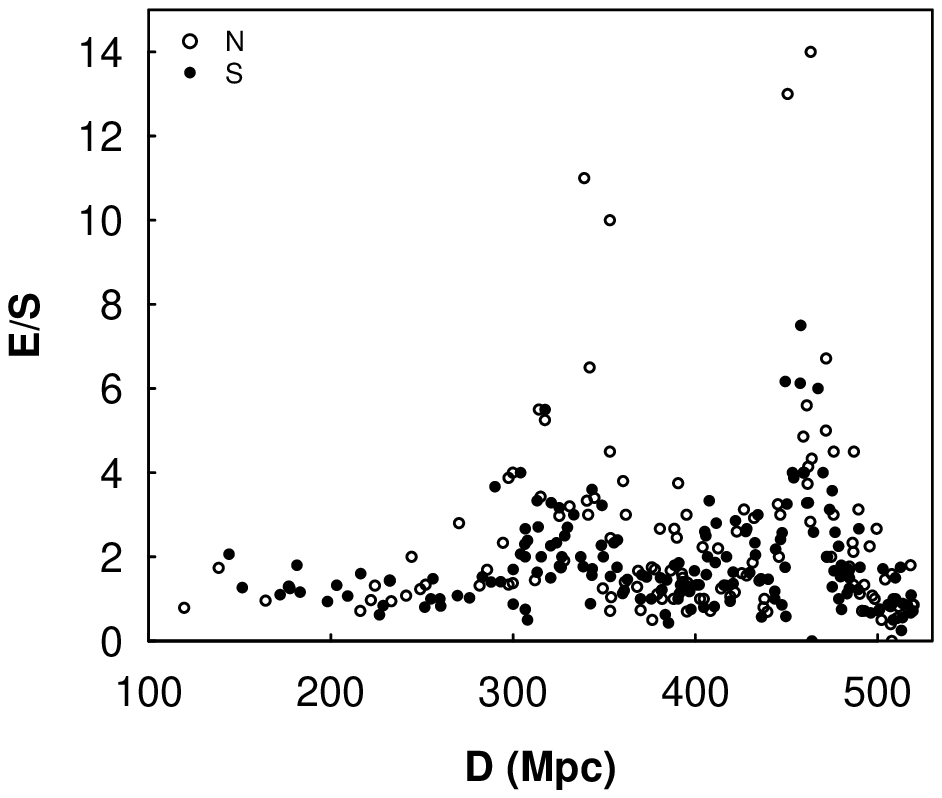}}
\resizebox{0.28\textwidth}{!}{\includegraphics*{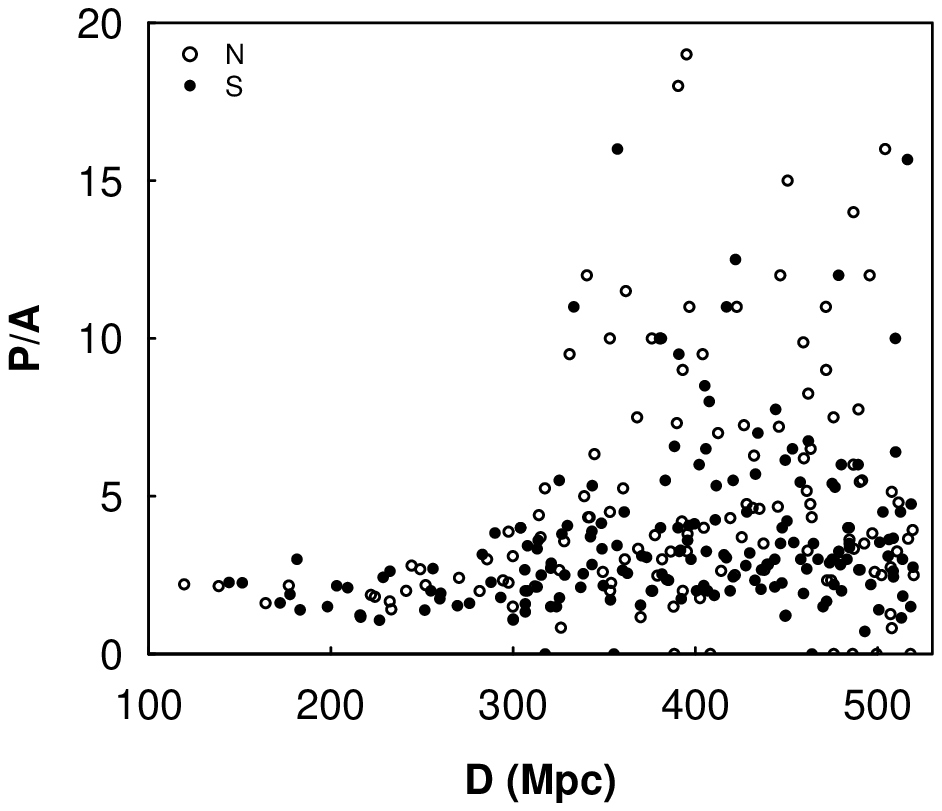}}
\hspace{2mm} \\
\raisebox{2pt}
{\resizebox{0.28\textwidth}{!}{\includegraphics*{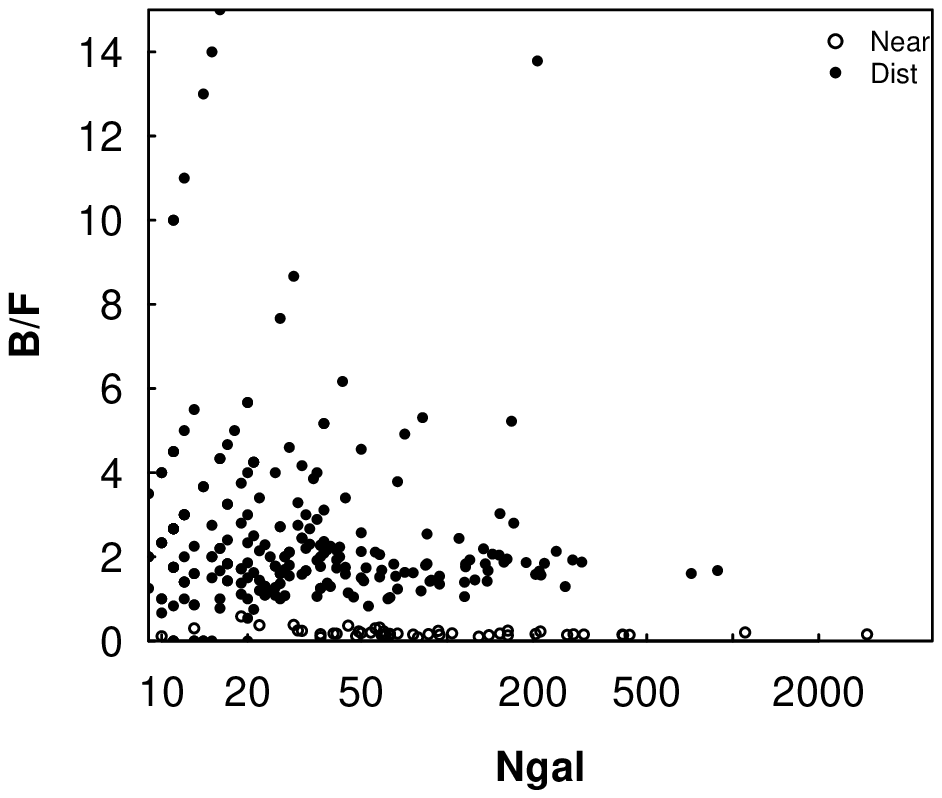}}}
{\resizebox{0.28\textwidth}{!}{\includegraphics*{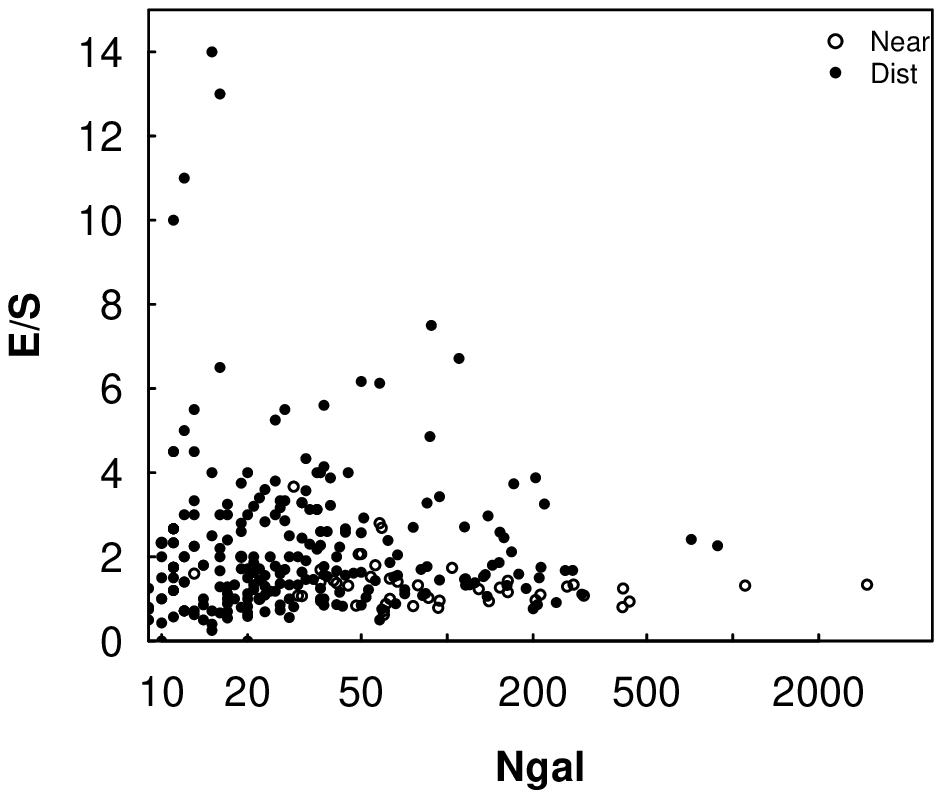}}}
{\resizebox{0.28\textwidth}{!}{\includegraphics*{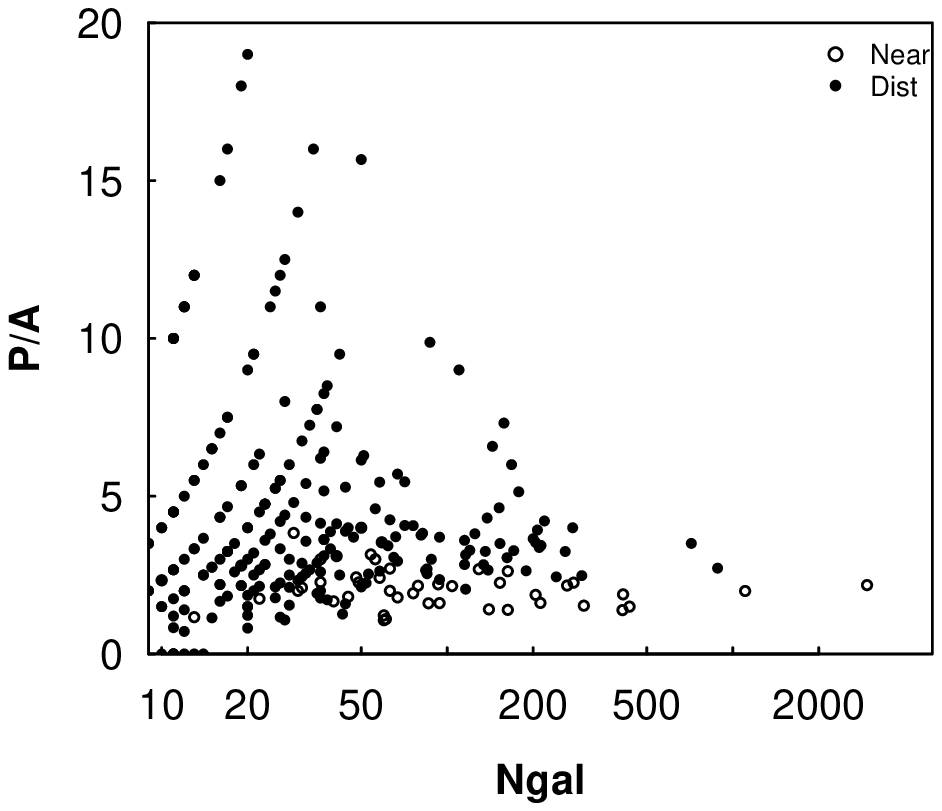}}}
\caption{The properties of galaxies in superclusters versus 
distances of superclusters (upper panels, 
  N --   Northern sky, S -- Southern sky) and
  versus the number of galaxies in superclusters (lower panels, N -- 
  nearby superclusters, D -- distant superclusters). Left: the ratio 
  of the numbers of bright and faint galaxies, middle:  the ratio 
  of the numbers of early and late type galaxies, right: the ratio 
  of the numbers of passive and actively star forming galaxies.
  divided by colour index $col$.
}
\label{fig:nsall}
\end{figure*}

Next we compare the properties of galaxies in individual superclusters. 
In Figure~\ref{fig:nsall} we plot
the ratio of the numbers of bright and faint galaxies $B/F$,
the ratio of the numbers of early and late type galaxies $E/S$,
and the ratio of the numbers of passive and actively star forming
galaxies  $P/A$ in superclusters with respect to the distance
of superclusters and to the number of galaxies.  

This Figure shows that the scatter of the ratio $B/F$ with distance and with the 
number of galaxies in superclusters (here this number is the number of galaxies 
in superclusters in volume-limited samples, not in full samples) is very small 
in the case of nearby samples. In distant samples the scatter of this ratio is 
larger, and increases with distance. This scatter is small for richer 
superclusters. Due to a higher luminosity cut-off in distant superclusters 
individual differences between poor distant superclusters affect this ratio more 
stongly than in nearby superclusters.

Comparison of the trends of the ratios of the numbers of early and late 
type galaxies 
$E/S$ with respect to the distance of superclusters and to the
number of galaxies in superclusters shows that this ratio
increases with distance (in Fig.~\ref{fig:1} we show the location of
superclusters with the ratio $E/S \geq 3$). Also, this ratio
is larger in poor superclusters; in some poor superclusters
the ratio $E/S$ is  over ten times larger than in superclusters in
average. 

We see similar trends, when we look at the ratio of passive and
actively star forming galaxies. 
In the case of distant superclusters the ratio $P/A$ has an even larger scatter than 
the ratio $E/S$.  In some poor superclusters
the number of passive galaxies is 15--20 times larger than the number of
actively star forming galaxies.  

This analysis shows that the properties of rich superclusters with at least 200 
member galaxies are rather homogeneous. This is one reason why we choose this 
richness limit to separate rich and poor superclusters. Additionally, the 
properties of nearby superclusters show only a small scatter; a larger scatter of 
the properties of distant superclusters is in accordance with our results in 
Paper II where we also saw a large scatter of the properties of poor distant 
superclusters. These are probably due to individual variations of the properties 
of poor superclusters. Also, some poor distant superclusters may be affected 
by selection effects (see Fig.~\ref{fig:2}).

\subsection{Density distribution in rich and poor superclusters}

{\scriptsize
\begin{table}[ht]
\caption{Environmental densities and KS test results
for galaxies of various populations in
rich and poor superclusters and for field galaxies. }
\begin{tabular}{lrrrrrr} 
ID     &    $N$     & 1Q&  Med &  3Q &     KS $D$   &     KS  $p$       \\
1      &     2      & 3 &   4  &   5 &     6        &     7          \\
R$_N$    &   7461  &  5.85 &  7.35  &   9.72  &         &                  \\
P$_N$    &   1652  &  4.94 &  5.38  &   6.10  &         &                  \\
FG$_N$    &  36949  &  0.99 &  1.96  &   2.80  &         &                  \\
R$_D$    &   5747  &  5.86 &  7.49  &   9.68  &         &                  \\
P$_D$    &   7031  &  5.40 &  6.33  &   7.71  &         &                  \\
FG$_D$    &  32162  &  1.41 &  2.34  &   3.39  &         &                  \\
     &          &       &        &          &         &                  \\
R$_{NB}$   &    1062  &  5.84&   7.32&    9.43    &         &                  \\
R$_{NF}$   &    6399  &  5.85&   7.34&    9.73   &  0.023 &      0.7   \\
P$_{NB}$   &     250  &  4.90&   5.37&    6.08   &         &                  \\
P$_{NF}$   &    1402  &  4.94&   5.38&    6.10   &  0.035 &       0.95   \\
FG$_{NB}$   &    4423  &  1.27&   2.05&    3.09   &         &                  \\
FG$_{NF}$   &   32526  &  0.95&   1.73&    2.75   &  0.015  &      0.000162    \\
R$_{DB}$   &    3741  &  5.91& 7.57 &   9.85   &         &                  \\
R$_{DF}$   &    2006  &  5.79& 7.31 &   9.43   &  0.042 &      0.018   \\
P$_{DB}$   &    4841  &  5.43& 6.42 &   7.81   &         &                  \\
P$_{DF}$   &    2190  &  5.34& 6.14 &   7.44   &  0.068 &       1.95e-6   \\
FG$_{DB}$   &   19563  &  1.52& 2.43 &   3.46   &         &                  \\
FG$_{DF}$   &   12599  &  1.23& 2.18 &   3.26   &  0.11  &      2.2e-16    \\
     &          &        &         &          &         &                  \\
R$_{NE}$   &   4044  & 6.00  & 7.57   &10.22    &         &                  \\
R$_{NS}$   &   2143  & 5.73  & 7.19   & 9.30    &  0.080 &      1.6e-10    \\
P$_{NE}$   &    888  & 4.97  & 5.44   & 6.15    &         &                  \\
P$_{NS}$   &    493  & 4.88  & 5.32   & 5.98    &  0.099  &      0.7e-4    \\
FG$_{NE}$   &   14262  & 1.18  & 1.99   & 3.04    &         &                  \\
FG$_{NS}$   &   14882  & 0.91  & 1.64   & 2.64    &  0.078 &      2.2e-16    \\
R$_{DE}$   &   3593  & 5.98  & 7.68   &9.92    &         &                  \\
R$_{DS}$   &   1213  & 5.60  & 6.94   & 9.01    &  0.07 &      3.32e-6    \\
P$_{DE}$   &   4361  & 5.45  & 6.42   & 7.81    &         &                  \\
P$_{DS}$   &   1481  & 5.32  & 6.17   & 7.49    &  0.063  &      7.8e-6    \\
FG$_{DE}$   &  15431  & 1.63  & 2.56   & 3.55    &         &                  \\
FG$_{DS}$   &   9411   & 1.27  & 2.17   & 3.26    &  0.11 &      2.2e-16    \\
      &          &        &         &          &         &                  \\
R$_{Nq}$  &    5192  &  5.94&   7.47&    9.98    &         &                  \\
R$_{NSF}$ &    2143  &  5.73&   7.08&    9.18    &  0.067 &      2.8e-6    \\
P$_{Nq}$  &    1129  &  4.95&   5.41&    6.14    &         &                  \\
P$_{NSF}$ &     493  &  4.87&   5.30&    5.96    &  0.087 &      0.011    \\
FG$_{Nq}$  &   21003  &  1.09&   1.91&    2.94   &         &                  \\
FG$_{NSF}$ &   14882  &  0.87&   1.59&    2.60    &  0.10 &      2.2e-16    \\
R$_{Dq}$  &    4401  &  5.91&   7.58&    9.82   &         &                  \\
R$_{DSF}$ &    1213  &  5.68&   7.15&    9.26   &  0.064 &      0.7e-4    \\
P$_{Dq}$  &    5346  &  5.42&   6.37&    7.75   &         &                  \\
P$_{DSF}$ &    1481  &  4.87&   5.30&    5.96   &  0.048 &      0.01    \\
FG$_{Dq}$  &   21101  &  1.54&   2.46&    3.49   &         &                  \\
FG$_{DSF}$ &    9411  &  1.25&   2.12&    3.21   &  0.10 &      2.2e-16    \\
     &          &       &        &          &         &                  \\
R$_{NP}$  &    4948  &  5.94&   7.49&  10.01  &         &                  \\
R$_{NA}$  &    2513  &  5.71&   7.08&   9.12  &  0.077 &      5.1e-9    \\
P$_{NP}$  &    1097  &  4.96&   5.43&   6.14  &         &                  \\
P$_{NA}$  &     555  &  4.87&   5.31&   5.98  &  0.086  &      0.009   \\
FG$_{NP}$  &   18764  &  1.12&   1.94&   3.00  &         &                  \\
FG$_{NA}$  &   18185  &  0.87&   1.58&   2.57  &  0.11 &       2.2e-16   \\
R$_{DP}$  &   4400   &  5.96&   7.62&   9.85  &         &                  \\
R$_{DA}$  &    1347  &  5.65&   7.15&   9.24  &  0.075 &      2.0e-5    \\
P$_{DP}$  &    5542  &  5.41&   6.35&   7.75  &         &                  \\
P$_{DA}$  &    1489  &  5.34&   6.23&   7.53  &  0.037  &      0.087   \\
FG$_{DP}$  &   22257  &  1.50&   2.44&   3.48  &         &                  \\
FG$_{DA}$  &    9905  &  1.21&   2.08&   3.16  &  0.10 &       2.2e-16   \\
\label{tab:3}
\end{tabular}
{

The Columns in the Table are as follows:

\noindent  1: Population ID. $R$ -- rich superclusters,
$P$ -- poor superclusters, $FG$ -- field galaxies;
$B$ -- bright galaxies ($M_{bj} \leq -20.0$),
$F$ -- faint  galaxies ($M_{bj} > -20.0$),
$E$ and $S$ -- early and late type galaxies,
$P$ and $A$ -- passive and actively star forming galaxies
(according to the spectral parameter $\eta$),
$q$ and \emph{SF} --  quiescent and actively star forming galaxies 
(according to the colour index $col$).
Index $N$ indicates the nearby population, $M_{bj}
\leq -18.4$ and $D \leq 300$\Mpc, index $D$ -- the distant population,  $M_{bj}
\leq -19.7$ and $D > 300$\Mpc.

\noindent  2: the number of galaxies in each population.

\noindent  3--5: the lower quartile, median and upper
quartile values of supercluster parameters.

\noindent  6 and 7: the Kolmogorov-Smirnov test results:
the maximum difference and the probability that the 
distributions of population parameters are taken
from the same parent distribution.
}

\end{table}
}

\begin{figure*}[ht]
\centering
\resizebox{0.45\textwidth}{!}{\includegraphics*{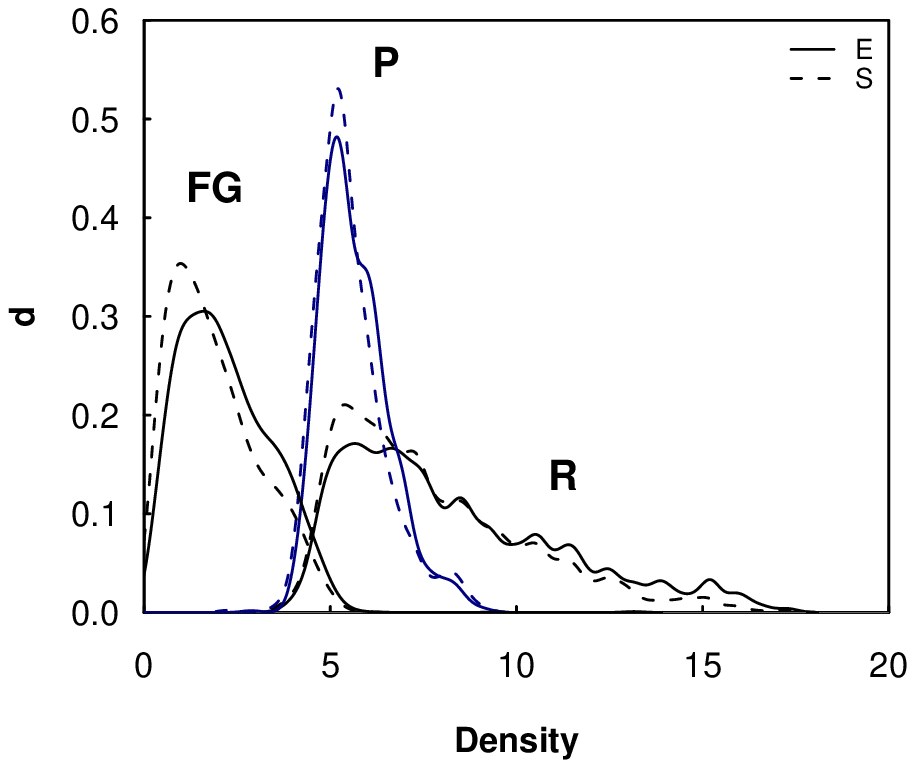}}
\hspace{2mm} 
\resizebox{0.45\textwidth}{!}{\includegraphics*{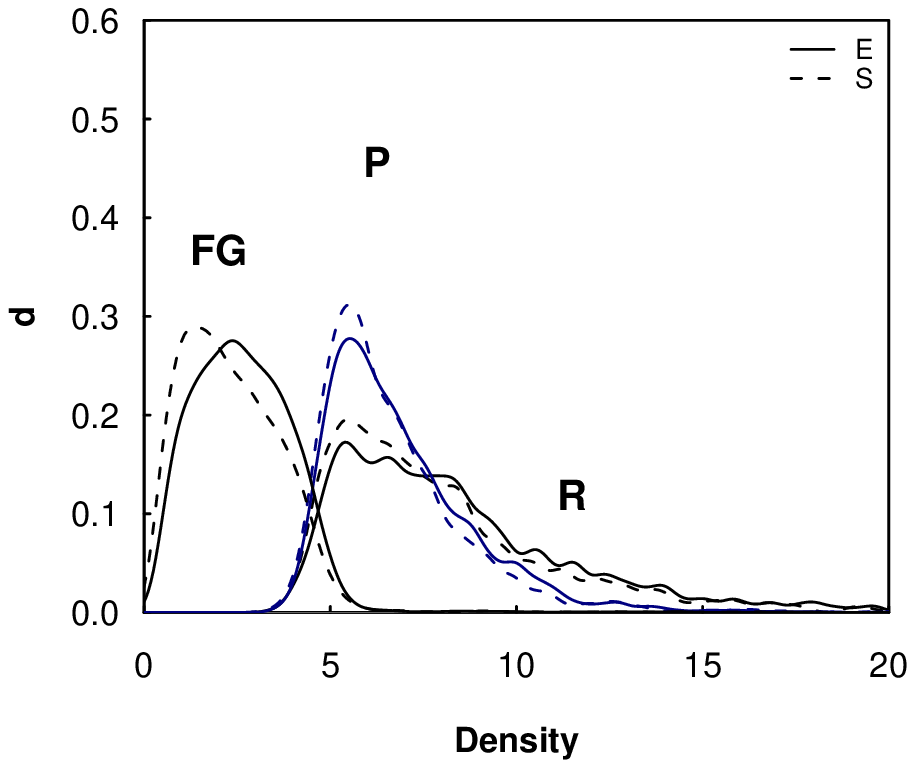}}
\caption{Environmental density distributions for 
  early and late type galaxies in rich (R) and poor (P) superclusters and in the
  field (FG). Left panel: $D \leq$ 300 \Mpc, right panel:
  $D \ge$ 300 \Mpc.  }
\label{fig:5}
\end{figure*}

\begin{figure*}[ht]
\centering
\resizebox{0.45\textwidth}{!}{\includegraphics*{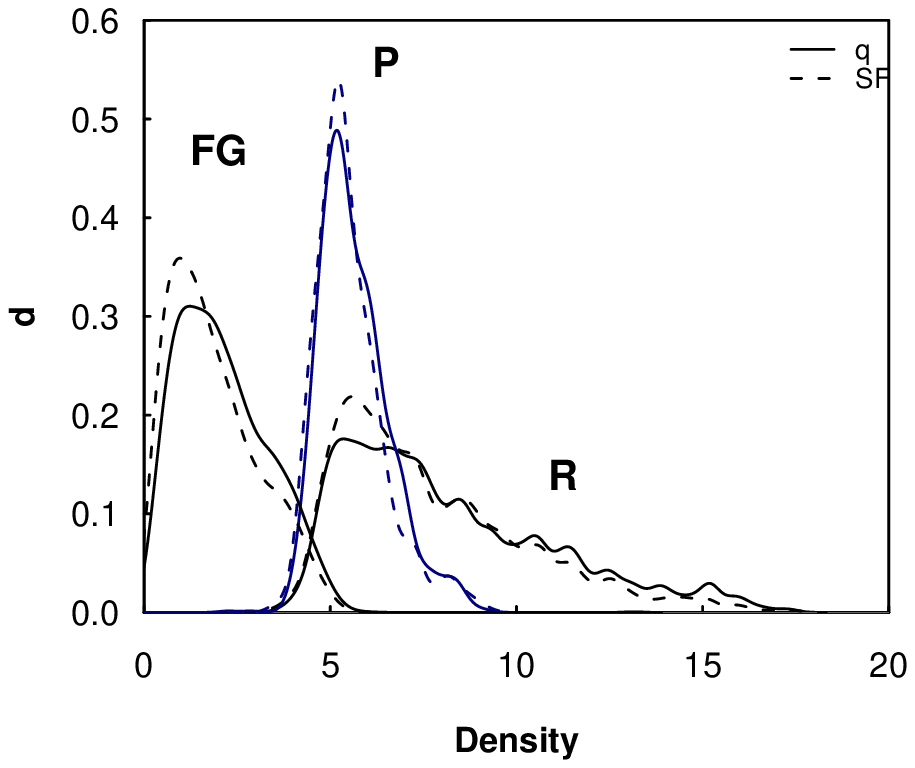}}
\hspace{2mm} 
\resizebox{0.45\textwidth}{!}{\includegraphics*{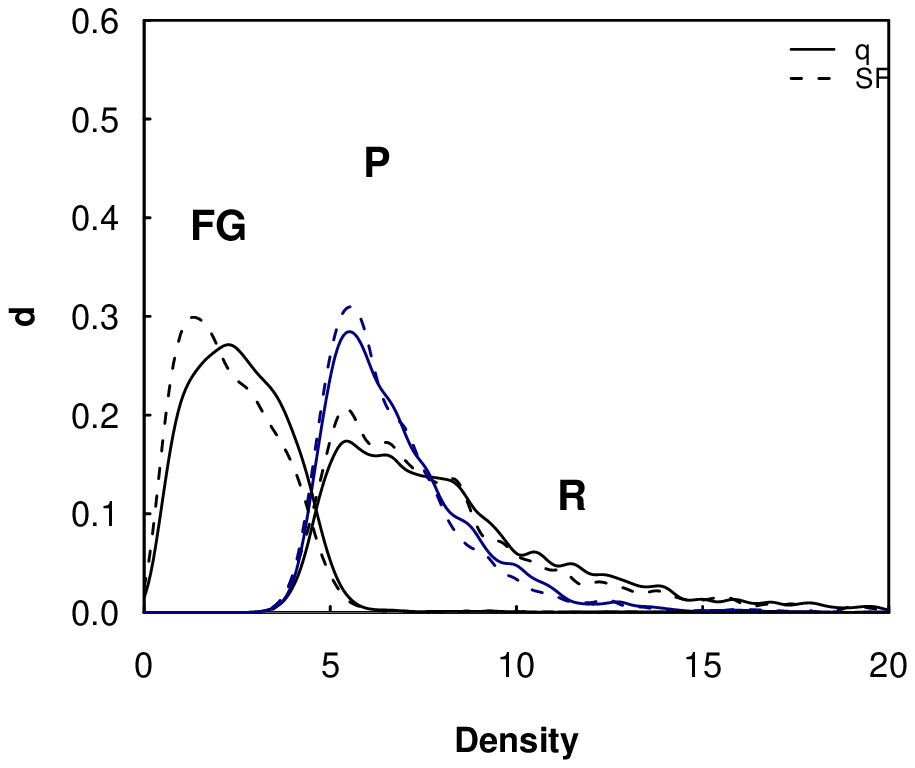}}
\caption{Environmental density distributions for
  quiescent and actively star forming  galaxies 
  (as determined by the spectral parameter $\eta$) in rich (R) and poor (P) 
  superclusters and in the
  field (FG). Left panel: $D \leq$ 300 \Mpc, right panel:
  $D \ge$ 300 \Mpc.  }
\label{fig:6}
\end{figure*}

\begin{figure*}[ht]
\centering
\resizebox{0.45\textwidth}{!}{\includegraphics*{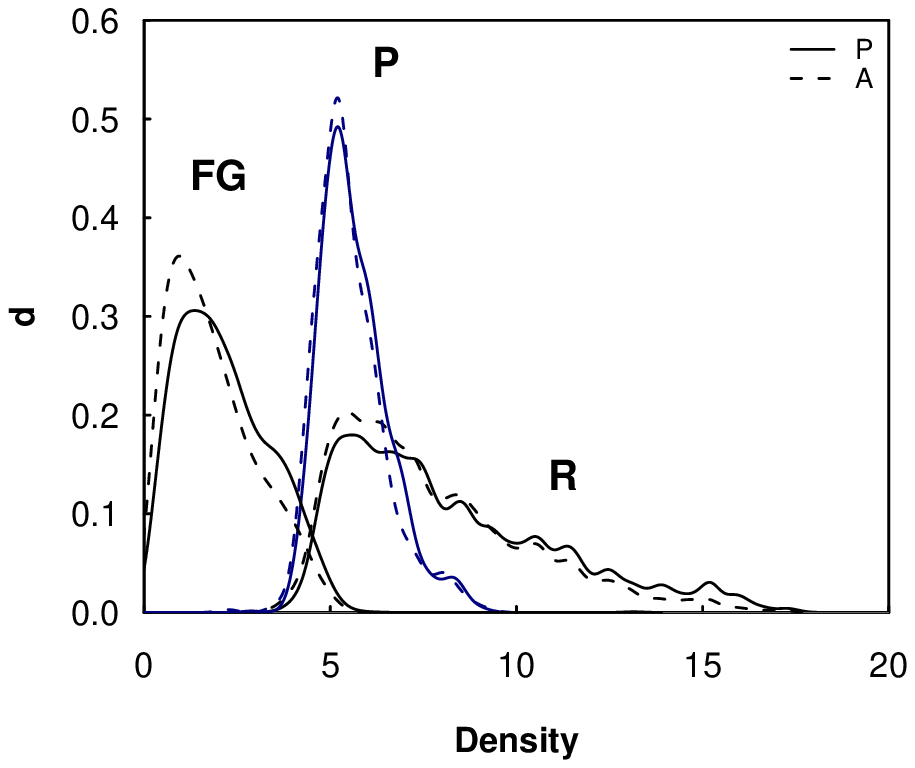}}
\hspace{2mm} 
\resizebox{0.43\textwidth}{!}{\includegraphics*{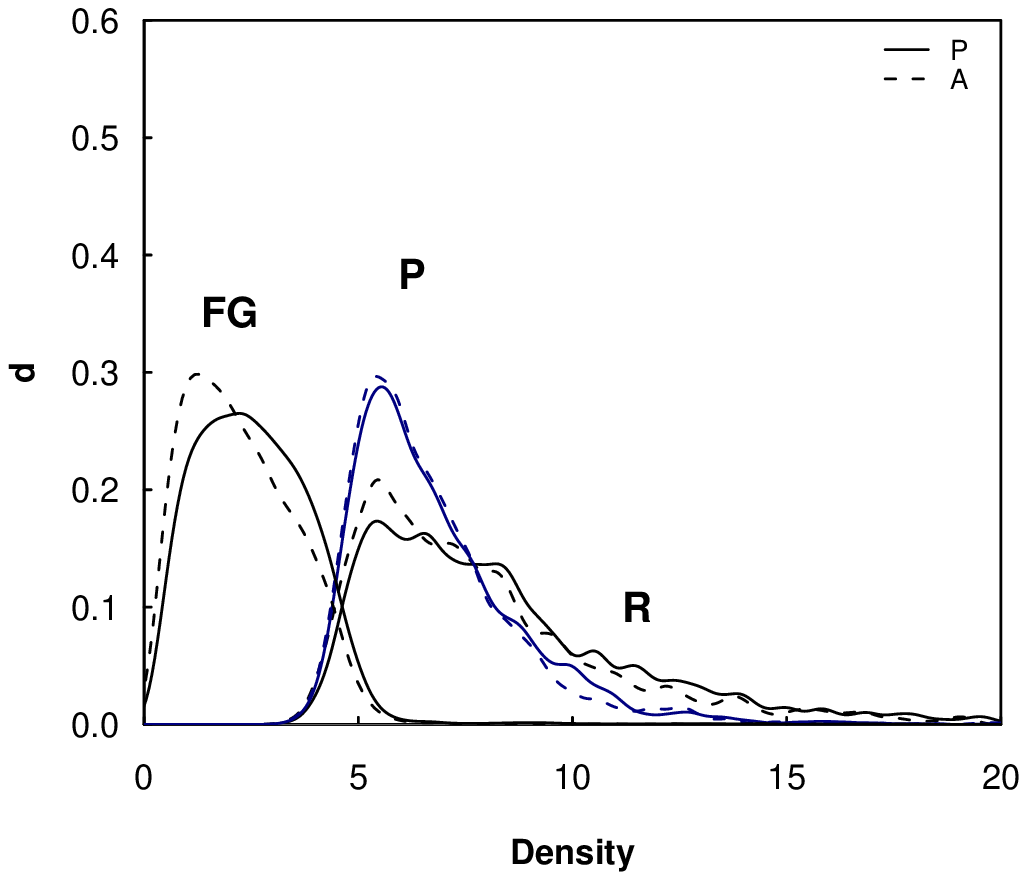}}
\caption{Environmental density distributions for  
 passive and actively star forming  galaxies 
  (as determined by the colour index $col$) in rich (R) and poor (P) superclusters 
  and in the field (FG). Left panel: $D \leq$ 300 \Mpc, right panel:
  $D \ge$ 300 \Mpc.  }
\label{fig:7}
\end{figure*}

Next we study how the properties of galaxies -- their luminosities, types
and activity -- depend on the large scale environment, defined as the value
of the density field at the location of galaxies
(environmental density).  We compare the distribution
of densities at the location of galaxies of each subsamples and for all
populations.

The results of this analysis are presented in Table~\ref{tab:3} and in
Figures~\ref{fig:5}-\ref{fig:7}.  In
Table~\ref{tab:3} we give the lower quantile, median and upper quantile values of
densities, as well as the results of the KS tests -- the
maximal differences between the density distributions $D$, and probabilities $p$, 
which show whether sample pairs may belong to the same parent sample.

Let us at first analyse the general distribution of densities 
in superclusters of various richness (see, e.g., Fig.~\ref{fig:5}).  
One remarkable feature seen in this Figure is the different
distribution of densities in rich and poor superclusters.  
Densities which
correspond to rich superclusters have a median value of 
$\delta \approx 7.5$, and the maximum densities are about 
$\delta \approx$ 17--20 (see also Paper II, where we showed that 
both mean and maximal densities in rich superclusters
are larger than in poor superclusters).  
These large densities show that rich superclusters contain high 
density cores. 
Densities in poor superclusters 
show a completely different distribution: they have a median value of
$\delta \approx 5.3 - 6.3$, and the maximum values densities 
are less than $\delta \approx 10$ for nearby poor superclusters and less than 
$\delta \approx 15$ in distant poor superclusters. Thus 
high density regions are absent in nearby 
poor superclusters; far away, even in some poor 
superclusters  high density cores are seen. It may be
possible that these poor superclusters are classified as poor due to selection
effects.

We remind that superclusters were defined as connected non-percolating 
systems with densities above a threshold density 4.6 in units of the 
mean luminosity density. We used an identical threshold density limit for 
superclusters of all richnesses. Then we divided superclusters by 
richness according to the number of galaxies in them, without using any 
additional condition about the values of the density field in 
supercluster regions. Thus these differences in density distributions 
reflect intrinsic properties of superclusters of various richness, and 
 not due to faults in the supercluster construction.

Next we analyse the distribution of densities at the location of
galaxies from different populations. 

Table~\ref{tab:3} shows the distribution of densities around bright and faint
galaxies. In nearby poor superclusters the differences between the density
distributions are not statistically significant (according to the KS test);
in nearby rich superclusters the probability that the density distribution 
around bright and faint galaxies are taken from the same parent distribution
is 0.7. In distant superclusters the differences are larger and their
statistical significance is higher. In the field the differences between
densities at the location of bright and faint galaxies are
statistically significant at a very high level.

Next we compare the density distributions at the location of 
galaxies of different type. Table~\ref{tab:3} 
shows that in all systems at the location of early type galaxies the values of the
density field  are larger than at the location of late type galaxies.  

A closer look at Fig.~\ref{fig:5} shows several interesting details. In 
both rich and poor superclusters at densities less than   
$\delta \approx 7$ there is an excess of late type galaxies. In rich 
superclusters at densities $\delta \geq 10$ there is seen an excess of early 
type galaxies. In distant poor superclusters there are also high density regions 
with an excess of early type galaxies and densities $\delta \geq 10$. 
As alreay noted, 
it may be possible that these poor superclusters are classified as poor due to 
selection effects.

Next we look at the density distributions at the location 
of quiescent and star forming galaxies, characterized by their 
spectral parameter $\eta$ (Table~\ref{tab:3}).
We see that the overall distributions of densities is 
rather similar to those for early and late type galaxies.
Passive galaxies are located at higher environmental  densities
than actively star forming galaxies.

Figure~\ref{fig:6} shows the density 
limits for regions where galaxies of different star formation 
rates dominate. In the case of nearby superclusters
 at densities less than  $\delta \approx 7$
there is an excess of star forming  galaxies in both rich and poor
superclusters. In regions with densities $\delta \geq 10$ 
passive galaxies dominate. These are the same density limits 
as those for early and late type galaxies. This is true also for 
distant superclusters.
 
Now let us study the distribution of the environmental 
densities for  galaxies divided into populations of 
passive (red) and active (blue) galaxies using colour information
(Table~\ref{tab:3}). 
As found before, passive galaxies have larger environmental densities
than actively star forming galaxies. 
We see in Fig.~\ref{fig:7}  
that also in this case the density limits for lower density regions
where star forming galaxies dominate, and for higher density regions
where passive galaxies dominate are the same as in the previous case.

The KS test confirms that the differences between environmental density
 distributions for galaxies of different properties are statistically
significant at very high levels (Table~\ref{tab:3}).  The differences between
the densities extend to the lowest amplitudes of the density field 
(for field galaxies).

To conclude, these Figures 
show correlation between the properties of galaxies 
and environmental density. There are also certain differences 
between rich and poor superclusters:
in rich superclusters there are 
higher density cores with $\delta > 10$
where early type, red, passive galaxies dominate,
and lower density regions $\delta < 7$ where there is an excess of late type,
blue, and actively star forming galaxies. In poor superclusters 
those high density regions are absent. This is another important
difference between rich and poor superclusters.

Among field galaxies we see also that at lower densities,
$\delta < 2.5$ late type, star forming galaxies dominate
while at higher densities, $\delta >  2.5$ there is an excess
of early type, passive galaxies.

\begin{figure*}[ht]
\centering
\resizebox{0.245\textwidth}{!}{\includegraphics*{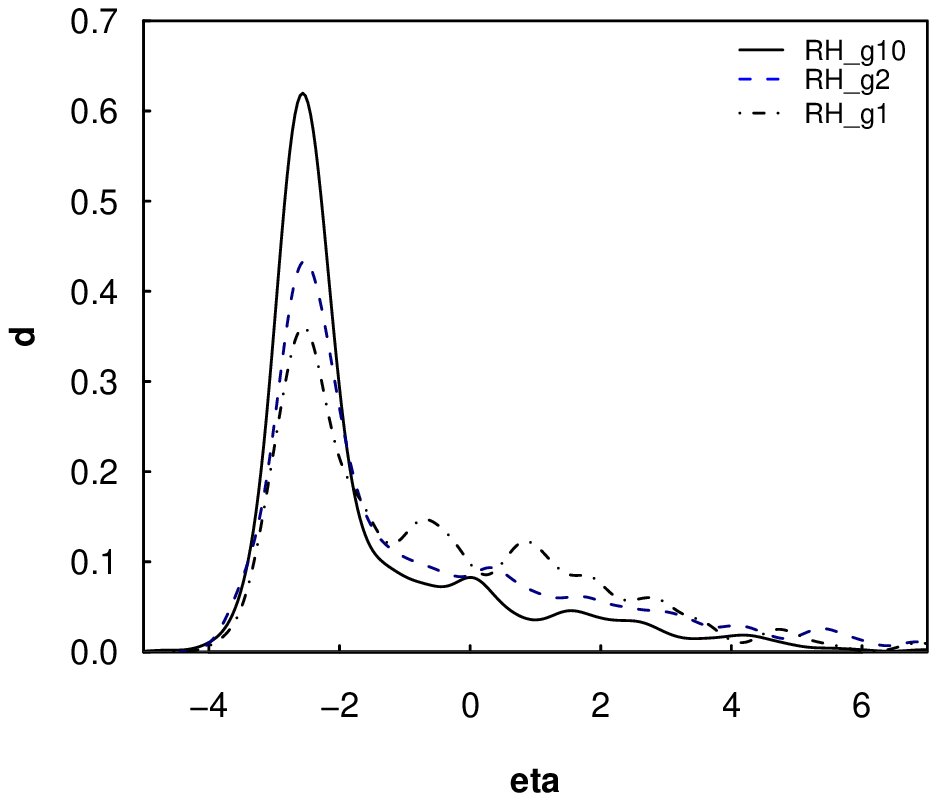}}
\resizebox{0.22\textwidth}{!}{\includegraphics*{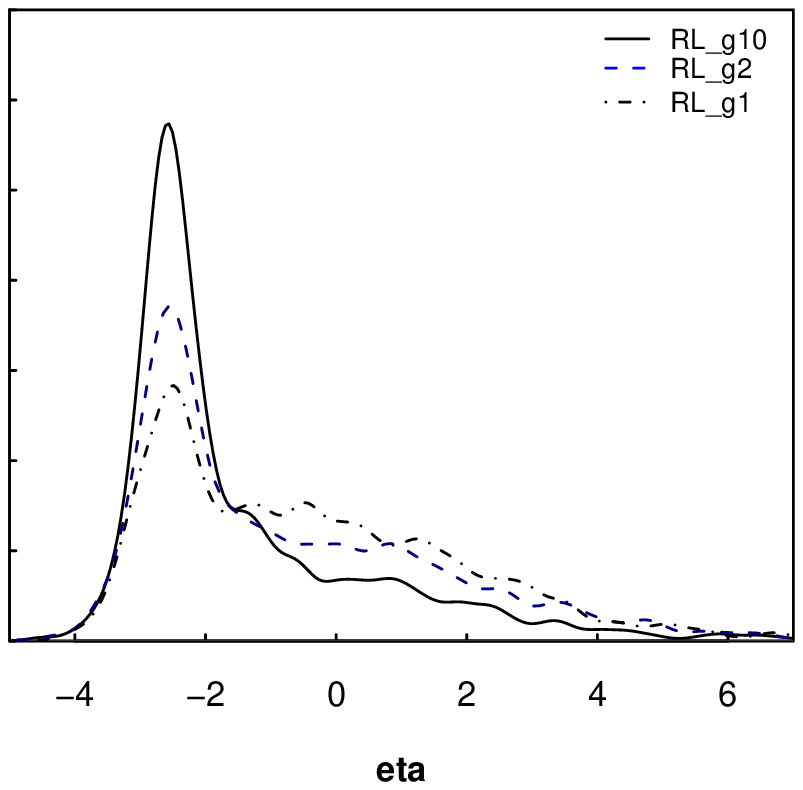}}
\resizebox{0.22\textwidth}{!}{\includegraphics*{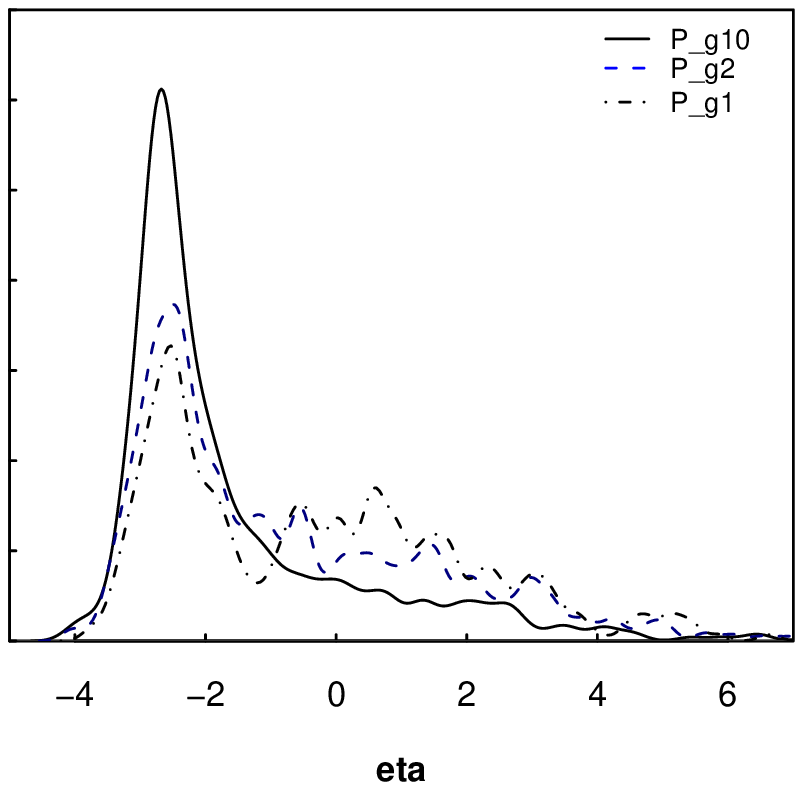}}
\resizebox{0.22\textwidth}{!}{\includegraphics*{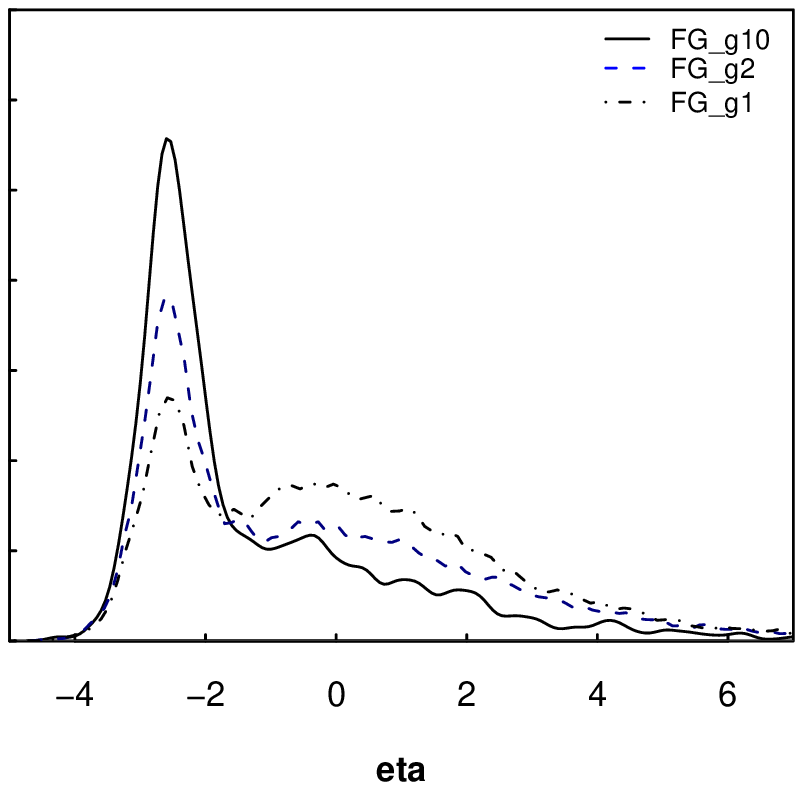}}
\\
\resizebox{0.25\textwidth}{!}{\includegraphics*{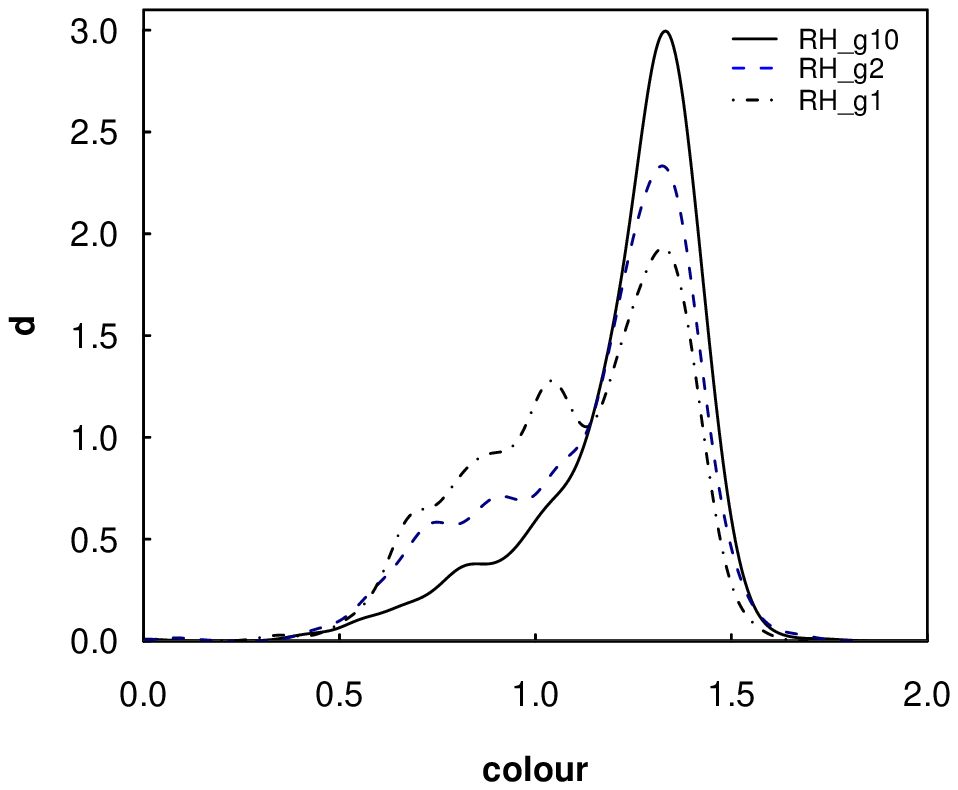}}
\resizebox{0.22\textwidth}{!}{\includegraphics*{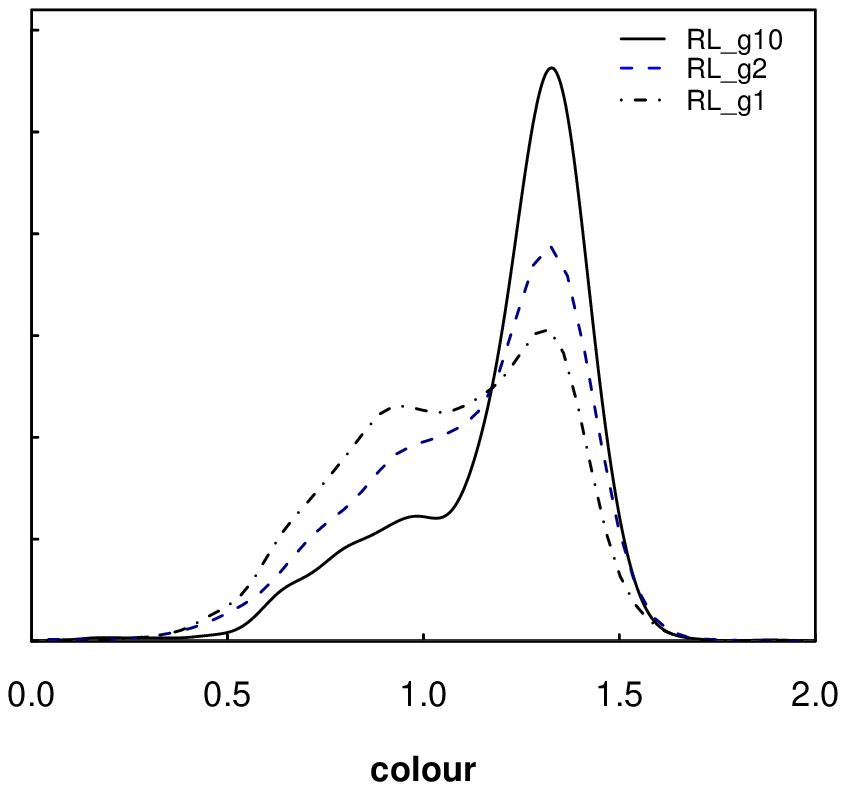}}
\resizebox{0.22\textwidth}{!}{\includegraphics*{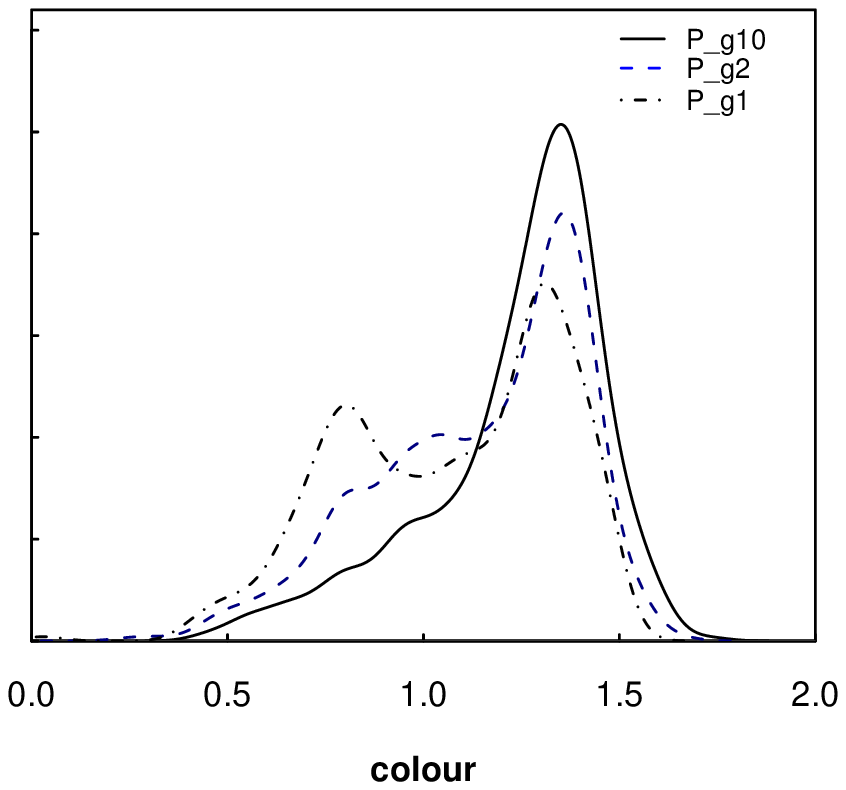}}
\resizebox{0.22\textwidth}{!}{\includegraphics*{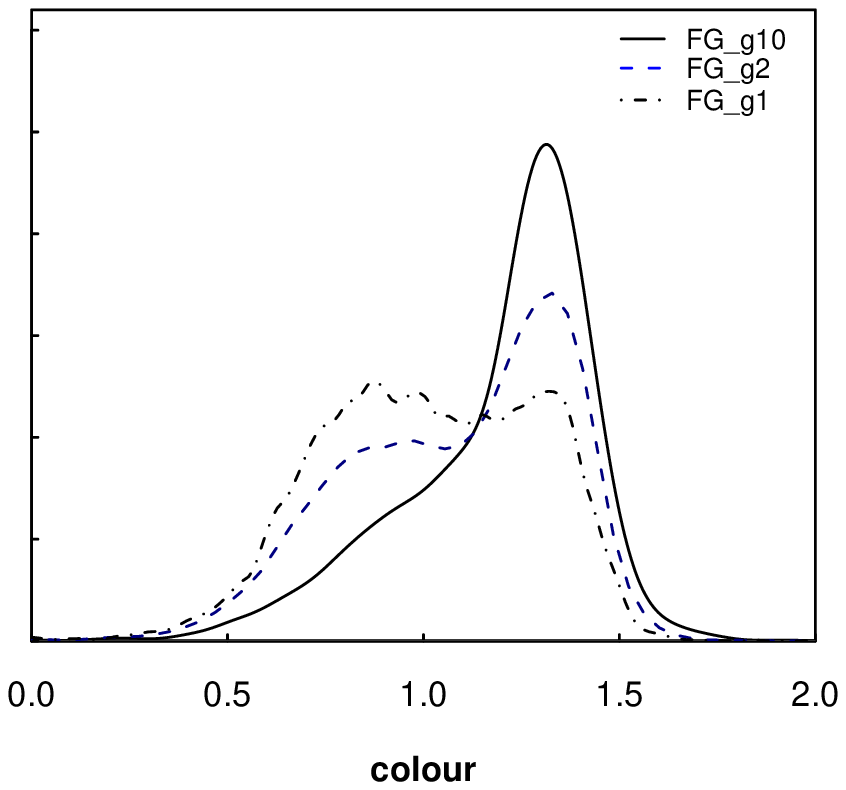}}
\caption{Distributions of the spectral parameter $\eta$ (upper panels)
and the colour index $col$ (lower panels) 
for galaxies in rich ($N_{gal} \geq 10$) and 
poor ($N_{gal} < 10$) groups, and for isolated galaxies.
From left to right: 
R$_{HD}$ -- high  density ($\delta \geq 10$) regions of rich superclusters,
R$_{LD}$ -- low   density ($\delta < 10$) regions of rich superclusters,
$P$ -- poor superclusters, $F$ -- field galaxies. 
}
\label{fig:rpigh}
\end{figure*}

\subsection{Properties 
of galaxies in rich and poor groups and of isolated galaxies  
in superclusters}
About 75\% of galaxies in superclusters and about 50\% of galaxies in the field 
belong to groups of galaxies of various richness. Next we compare the properties 
of galaxies in rich and poor groups and of isolated galaxies in various 
environments (in high and low density regions of rich superclusters, in poor 
superclusters and in the field). In this analysis we divide groups by their 
richness as follows:  rich ($N_{gal} \geq 10$) groups and clusters (we denote 
this sample as $Gr_{10}$) and poor groups ($N_{gal} < 10$, $Gr_2$).  Those 
galaxies in  superclusters and in the field which do not belong to groups form a 
sample of isolated galaxies ($I$). In other words, we analyse how both the 
small (group) scale and large (supercluster) scale environments influence the 
properties of galaxies. In this analysis we use the data about nearby 
superclusters only, as these data are more reliable. The results of this analysis 
are presented in Table~\ref{tab:4} and in Figure~\ref{fig:rpigh}. In 
Table~\ref{tab:5} we give the results of the Kolmogorov-Smirnov test showing 
the probability that the distributions of the spectral parameters $\eta$ and the colour 
index $col$ are drawn from the same parent sample.

Fig.~\ref{fig:rpigh} and Table~\ref{tab:4} show additional interesting
features. Let us analyse, for example, the presence of passive and
actively star forming galaxies according to the colour information in groups
of various richness in superclusters and in the field.
The fraction of passive (red) galaxies is largest in rich groups
in the high density cores of rich superclusters.
Even in poor groups in these high density regions the fraction of
passive galaxies is larger than this fraction in poor groups in 
poor superclusters and in the field. And finally, the fraction
of passive galaxies among galaxies which do not belong to groups,
but are located in superclusters, is 1.5 times higher than among
isolated galaxies in the field. This shows that  
star formation in  galaxies in high density cores of superclusters
is supressed even for isolated galaxies, not only in rich groups
and clusters in these regions.

The fraction of star forming galaxies in low density regions 
of rich superclusters and in poor superclusters is similar.
In the field, star forming galaxies are as abundant as isolated
supercluster galaxies, while passive galaxies dominate in groups.

Fig.~\ref{fig:rpigh} shows also a shift to blue colours 
when we compare the distribution of the colour index  $col$
for isolated galaxies in superclusters and in the field.
Therefore the differences between colour distributions berween
these populations are even larger than the ratio $P/A$ shows.

We see similar trends when we study the fractions
of quiescent and actively star forming galaxies, according to  spectral 
information, in groups of various richness in superclusters and in the field.
In high density regions of rich superclusters the fraction of passive
galaxies is the largest. Even among isolated galaxies
in these regions the ratio of the numbers of quiescent and actively 
star forming galaxies is comparable to that of in poor groups in 
less dense enviromnents, in low density regions of rich superclusters
and in poor superclusters.

The ratio of the numbers of early and late type 
galaxies shows similar trends: in high density cores early type
galaxies dominate both in rich and in poor groups; 
even among isolated galaxies in these regions
late and early type galaxies are almost equally present
while in poor superclusters and in low density regions of rich 
superclusters late type galaxies dominate among isolated galaxies.

These results show that both local (group/cluster)
environments and global (supercluster) environments are  important
in forming galaxy morphologies and star formation activity.

A weak dependence of the ratio of the numbers of bright and faint galaxies
on the environment, found  in the present 
analysis, is caused by the fact that 
we used the data about relatively bright galaxies only. Among isolated
galaxies the fraction of bright galaxies is smaller, thus the differences
between the properties of galaxies in groups and isolated galaxies 
come at least partly from the luminosity difference.
However, in superclusters of different richness
the ratio $B/F$ for galaxies in groups and for isolated galaxies 
is similar, thus the differences between the properties
of galaxies in rich and poor superclusters are not due to different 
luminosities of galaxies. 

{\scriptsize
\begin{table}[ht]
\caption{Properties of galaxies in groups of various richness
in superclusters. }
\begin{tabular}{lrrrrrr} 
\hline 
\\ 
ID     &    R$_H$   & R$_L$ &  P &  FG   \\
1     &    2   & 3 &  4 &  5   \\
\hline 
\\ 
$N_{gal}$ &&&& \\
$Gr_{10}$&   878   & 1664  &   501  &   1788     \\
$Gr_2$   &   536   & 2528  &   769  &  15981     \\
$I$      &   285   & 1570  &   382  &  19180     \\
\\
$E/S$ &&&& \\
$Gr_{10}$&  2.52   & 2.00  &  2.46  &  1.67      \\
$Gr_2$   &  1.35   & 1.03  &  1.04  &  0.82       \\
$I$      &  0.93   & 0.70  &  0.69  &  0.49       \\
\\
$q/$\emph{SF} &&&& \\
$Gr_{10}$&  4.87   & 4.04  &  4.69  &  3.47      \\
$Gr_2$   &  2.44   & 2.01  &  2.09  &  1.61       \\
$I$      &  1.99   & 1.59  &  1.30  &  1.17       \\
\\
$P/A$ &&&& \\
$Gr_{10}$&  4.23   & 3.37 &  3.91  &   2.89   \\
$Gr_2$   &  2.21   & 1.72 &  1.73  &   1.29   \\
$I$      &  1.28   & 1.12 &  1.23  &   0.78   \\
\\
\hline 
$B/F$ &&&& \\
$Gr_{10}$&  0.16   &  0.19 &  0.20  &  0.20       \\
$Gr_2$   &  0.18   &  0.18 &  0.19  &  0.16       \\
$I$      &  0.10   &  0.13 &  0.11  &  0.11       \\
\\
\hline 
\label{tab:4}
\end{tabular}
{

The Columns in the Table are as follows:

\noindent  1: Group membership: $Gr_{10}$ -- galaxies in rich ($N_{gal} \geq 10$)
groups, $Gr_2$ -- galaxies in poor ($N_{gal} < 10$) groups, $I$ -- isolated 
galaxies (i.e. galaxies which do not belong to groups);

\noindent  2--5: Populations: 
R$_H$ -- high density regions ($\delta \geq 10$) of 
rich superclusters, R$_L$ -- low density regions ($\delta < 10$) of 
rich superclusters,
P -- poor superclusters, FG -- field galaxies;

$E/S$ -- the ratio of the numbers of early and late type galaxies,
$q/$\emph{SF}
-- the ratio of the numbers of quiescent and actively star forming  galaxies 
(according to the spectral parameter $\eta$), $P/A$ --  the ratio of the numbers 
of passive and actively star forming galaxies  (according to the colour 
index $col$), $B/F$ -- the ratio of the numbers of the bright ($M_{bj}
\leq -20.0$) and  faint ($M_{bj} > -20.0$) galaxies.

}

\end{table}
}

{\scriptsize
\begin{table}[ht]
\caption{Properties of galaxies in groups of various richness
in superclusters. The Kolmogorov-Smirnov test results.}
\begin{tabular}{lrrr} 
\hline 
ID1       &    ID2   &    $D$     &  $P$   \\
1       &    2   &    3     &  4   \\
\hline 
$col$      &       &         &     \\
\hline 
RH$_{Gr10}$ & RH$_{Gr2}$    &   0.1444   &    1.870e-06     \\
RH$_{Gr2}$    & RH$_I$       &   0.1275   &    0.004715      \\
   \\                       
RH$_{Gr10}$ & RL$_{Gr10}$ &   0.0539   &    0.07127       \\
RH$_{Gr2}$    & RL$_{Gr2}$  &   0.0801   &    0.00684       \\
RH$_I$       & RL$_I$       &   0.0832   &    0.07094       \\
    \\                      
RL$_{Gr10}$ & RL$_{Gr2}$  &   0.1692   &   $<$ 2.2e-16    \\
RL$_{Gr2}$  & RL$_I$       &   0.1249   &    1.49e-13      \\
   \\
RL$_{Gr10}$ & P$_{Gr10}$  &   0.1034   &    0.000529      \\
RL$_{Gr2}$  & P$_{Gr2}$   &   0.0686   &    0.007783      \\
RL$_I$       & P$_I$        &   0.0724   &    0.08008       \\
    \\
P$_{Gr10}$ & P$_{Gr2}$    &   0.1897   &    6.639e-10     \\
P$_{Gr2}$  & P$_I$         &   0.1245   &    0.0007295     \\
   \\
P$_{Gr10}$ & FG$_{Gr10}$  &   0.1105   &    0.0001426     \\
P$_{Gr2}$  & FG$_{Gr2}$   &   0.1064   &    1.228e-07     \\
P$_I$       & FG$_I$        &   0.1379   &    1.304e-06     \\
   \\
FG$_{Gr10}$ & FG$_{Gr2}$  &   0.1909   &   $<$ 2.2e-16    \\          
FG$_{Gr2}$  & FG$_I$       &   0.1353   &   $<$ 2.2e-16    \\
\hline 
$\eta$      &       &         &     \\
\hline 
RH$_{Gr10}$ & RH$_{Gr2}$     &   0.1658       &     2.801e-08     \\
RH$_{Gr2}$    & RH$_I$        &   0.112        &     0.02062       \\
   \\
RH$_{Gr10}$ & RL$_{Gr10}$  &   0.069        &     0.008986      \\
RH$_{Gr2}$    & RL$_{Gr2}$   &   0.0799       &     0.007649      \\
RH$_I$       & RL$_I$        &   0.088        &     0.05248       \\
       \\
RL$_{Gr10}$ & RL$_{Gr2}$   &   0.1682       &   $<$ 2.2e-16     \\
RL$_{Gr2}$  & RL$_I$        &   0.1029       &     3.799e-09     \\
         \\
RL$_{Gr10}$ & P$_{Gr10}$   &   0.0802       &     0.01511       \\
RL$_{Gr2}$  & P$_{Gr2}$    &   0.032        &     0.5936        \\
RL$_I$       & P$_I$         &   0.0696       &     0.1084        \\
        \\
P$_{Gr10}$  & P$_{Gr2}$    &   0.2018       &     5.521e-11     \\
P$_{Gr2}$   & P$_I$         &   0.1478       &     3.589e-05     \\
     \\
P$_{Gr10}$  & FG$_{Gr10}$  &   0.1098       &     0.0001757     \\
P$_{Gr2}$   & FG$_{Gr2}$   &   0.0867       &     4.108e-05     \\
P$_I$        & FG$_I$        &   0.0905       &     0.004841      \\
    \\
FG$_{Gr10}$ & FG$_{Gr2}$   &   0.1796       &   $<$ 2.2e-16     \\
FG$_{Gr2}$  & FG$_I$        &   0.121        &   $<$ 2.2e-16     \\
     \\
RH$_{Gr10}$ & P$_{Gr10}$   &   0.0666       &     0.1217        \\
RH$_{Gr2}$    & P$_{Gr2}$    &   0.0719       &     0.08098       \\
RH$_I$       & P$_I$         &   0.1187       &     0.02229       \\
\hline 
\label{tab:5}
\end{tabular}
{

The Columns in the Table are as follows:

\noindent  1--2: Sample ID. $Gr_{10}$ -- galaxies in rich ($N_{gal} \geq 10$)
groups, $Gr_2$ -- galaxies in poor ($N_{gal} < 10$) groups, $I$ -- isolated 
galaxies (i.e. galaxies which do not belong to groups);
RH -- high density cores of rich superclusters,
RL -- low density regions of rich superclusters,
P -- poor superclusters, FG -- field galaxies. 

\noindent  3--4: the Kolmogorov-Smirnov test results:
the maximum difference and the probability that the 
distributions of population parameters are taken
from the same parent distribution.
}

\end{table}
}

\subsection{Luminosities of supercluster main galaxies}

\begin{figure*}[ht]
\centering
\resizebox{.46\textwidth}{!}{\includegraphics*{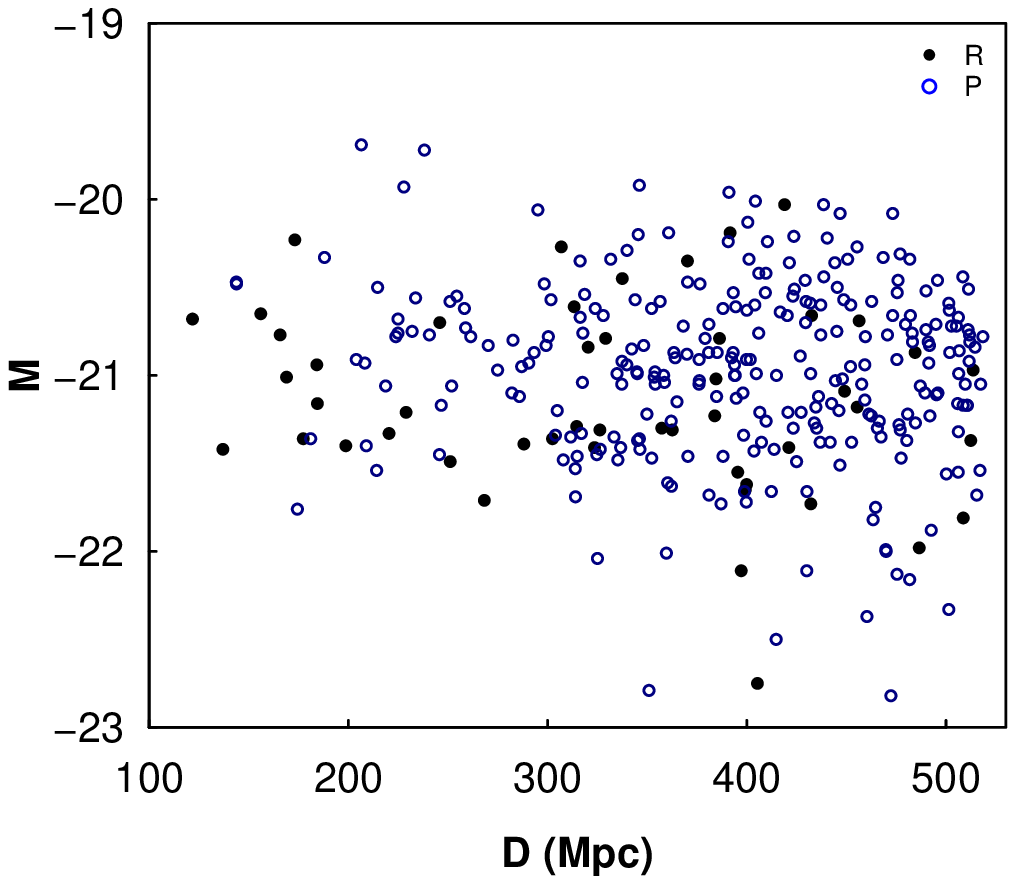}}
\resizebox{.48\textwidth}{!}{\includegraphics*{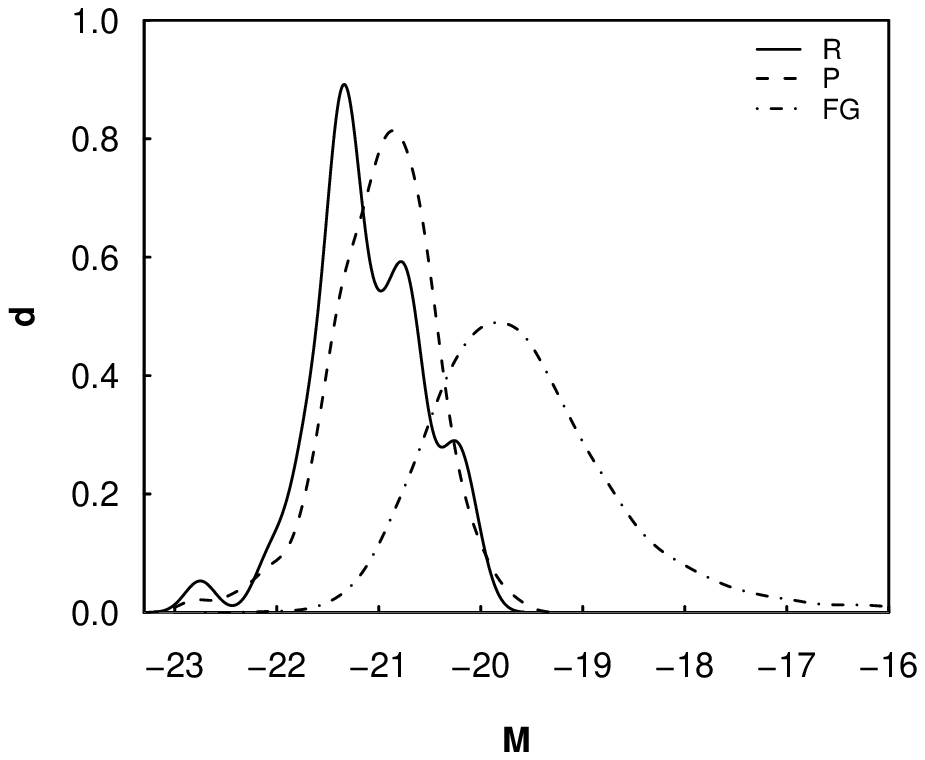}}
\\
\caption{Left panel: the luminosity of the main galaxy of 
  rich (R) and poor (P) superclusters vs. the
  distance of supercluster.  Right panel: the distributions of the
  luminosities of main galaxies of rich (R) and poor (P)
  superclusters and of groups in the field (FG).  }
\label{fig:11}
\end{figure*}

We determined for each supercluster the main group and its main galaxy as
described in Paper I: the most luminous cluster in the vicinity of the highest
density peak in a supercluster is considered as the main cluster and its
brightest galaxy -- the main galaxy of the supercluster.  When determining main
galaxies we used an automated search routine, without using supplementary
information on the morphological type, colour etc.

Fig. \ref{fig:11} (left panel) shows the luminosities of main galaxies of
superclusters at various distances from the observer.  Since almost all main
galaxies have a luminosity higher than the 2dF survey limit the trend with
distance is weak.  Note the narrow range of luminosities of main galaxies.

Fig.~\ref{fig:11} (right panel) shows the distributions of luminosities of
supercluster main galaxies.  This Figure shows that main galaxies of rich
superclusters have larger luminosities than those of poor superclusters. The
median luminosities of main galaxies of rich and poor superclusters
and of groups in the field are, correspondingly, $-$21.2, $-$20.9 and $-$19.7
magnitudes.  The Kolmogorov-Smirnov test shows that the probability that the
luminosities of main galaxies of rich and poor superclusters are taken from
the same parent distribution is less than $0.05$, the probability that the
luminosities of main galaxies of superclusters and of groups in the field are
taken from the same parent distribution is less than $10^{-16}$.

\section{Discussion and conclusions}

A detailed study of luminosity functions of galaxies from the 2dF survey in
regions of different density of the large-scale environment was made by Croton
et al.  (\cite{cr05}), and in clusters by De Propris et al. (\cite{depr03}). 
De Propris et al. (\cite{depr03}) found that the luminosity functions of early 
type galaxies in clusters are brighter and steeper than those in the field, and
that clustering of passive galaxies is stronger than clustering of actively star
forming galaxies (Madgwick et al. \cite{ma03b}). Using densities smoothed on a
scale of 8~\Mpc\, Croton et al.  (\cite{cr05}) divided the volume under study
into 7 regions of various environmental densities, from extreme voids to
cluster populations and found that the brightest galaxies in voids are
approximately 5 times fainter than those in clusters. Even larger differences
between luminosities of galaxies in high and low density regions were found in
Einasto et al. (\cite{e05b}). In the present paper we showed, in accordance to
these results, that the luminosity-density dependence in important
for galaxies of all types both 
in superclusters and among field galaxies at all densities.

Several recent studies address the problem whether the properties of 
galaxies, their star formation activity in particular, correlates with the 
local and/or global environment of galaxies, defined, for example,
as the clustercentric distance. The well-known morphology-density 
relation is an example of such correlation (Dressler (\cite {d80}).
One question asked in these studies is whether there exists a critical
density so that star formation is supressed in all
groups/clusters where the density exceeds this critical value.
Lewis et al. (\cite{lew02}) used the data about star-forming galaxies 
in the 2dFGRS to show that there exists a correlation between 
the star formation activity and the local galaxy density which holds for galaxies
at distances at least two virial radii from the cluster/group centre.

Balogh et al. (\cite{balogh04}) compared the populations of
star-forming and quiescent galaxies 
in small (1.1~\Mpc) and large scales (5.5~\Mpc)
 and showed that the relative numbers of these 
galaxies depend both on the local and global environments. 
Even low density environments 
contain a large fraction of non-star-forming galaxies.
They concluded that the galaxy population must be only indirectly
related to their present-day environment. Possible
physical mechanisms must have been more effective in the past, and perhaps
affeced the star formation rate in very short (less than 1 Gyr) timescales,
like starbursts induced by galaxy interactions in close pairs of galaxies
(see Balogh et al. \cite{balogh04} and references therein).

Gray et al. (\cite{gray04}) found, using data about the supercluster A901/902,
strong evidence that the highest density regions in clusters 
are populated
mostly with quiescent galaxies, while star forming galaxies dominate
in outer/lower density regions of clusters.
Similarly, Haines et al. (\cite{hai06}) demonstrated that the colours of
galaxies in the core region of the Shapley supercluster
depend on environment, redder galaxies being located in cluster cores.
They also found large concentrations of faint blue 
galaxies between clusters. 

Our results are in accordance with those;in addition we showed 
that even at supercluster scales, the properties of galaxies and their
environmental densities are correlated. In high density cores of rich superclusters
the fraction of quiescent (red) galaxies is higher than
this fraction in lower density regions even for those galaxies which 
do not belong to groups or clusters.

Porter and Raychaudhury (\cite{pr05}) investigated the star formation rate in
groups of galaxies from the Pisces-Cetus supercluster, according to their
spectral index $\eta$. They concluded that galaxies in rich clusters have
lower star formation rates than galaxies in poor groups.  This is in
accordance with our results, showing that galaxies from a higher density
environment have lower star formation rates than galaxies from a lower density
environment.

Our previous analysis in Paper II demonstrated that geometrical properties
of rich and poor superclusters are different, rich superclusters have larger
sizes, their shapes and compactness differ from those of poor superclusters.
The mean density of superclusters  increases gradually with increasing total
luminosity or richness of superclusters. This demonstrates
that rich superclusters are physical systems with properties different from
those of poor systems, they do not represent just percolations of several 
loose systems.

Our present study reveals additional differences between rich and poor 
superclusters. Rich superclusters contain high density cores which are absent in 
poor supercluters. The fraction of early type and passive galaxies in groups and 
clusters and among isolated galaxies in high density cores of rich superclusters 
is higher than in groups in poor superclusters. 

We shall analyse the detailed properties of high density regions in rich 
superclusters in another study (Einasto et al., in preparation).

Hilton et al. (\cite{hil05}) found that the fraction of early spectral type
galaxies is significantly higher in clusters with a high X-ray flux. Many of
these clusters belong to rich superclusters (Einasto et al. \cite{e2001},
Belsole et al. \cite{bel04}), so this result is in accordance with the 
present paper.

Our analysis  shows that main galaxies of superclusters form a specific 
class of galaxies with a very limited range of luminosities. Main 
galaxies of rich superclusters are more luminous than main galaxies of 
poor superclusters. A similar conclusion has been reached by Einasto \& 
Einasto (\cite{e87}) using data on nearby superclusters. This is in 
accordance with the result of De Propris et al. (\cite{depr03}) that 
there is an excess of very bright galaxies in cores of clusters. Main 
galaxies of superclusters are formed by multiple merger processes, as 
indicated by direct observations and numerical simulations (see Laine et 
al. \cite{laine03} and Gao et al. \cite{gao05}). It has been known 
already for a long time that first-ranked cluster galaxies have a small 
dispersion of absolute magnitudes (Hubble \& Humason \cite{hubble31}, 
Hubble \cite{hubble36}, Sandage \cite{sandage76}).  More recent studies 
by Postman \& Lauer (\cite{postman95}) and Laine et al. (\cite{laine03}) 
have shown that absolute magnitudes of brightest cluster galaxies have a 
scatter of about 0.24 - 0.33 mag. The scatter of luminosities of 
supercluster main galaxies is larger than the scatter of brightest 
cluster galaxies.  There may be several reasons for this.  One 
possibility is that we have found the main group and its main galaxy by 
an automated search routine, and supplementary information on the 
morphological type, colour etc. has not been used. For this reason our 
sample of main galaxies is probably not as homogeneous as samples of 
first-ranked cluster galaxies investigated by Hubble, Sandage, Postman 
and others.

Our analysis shows a large scatter of the properties of galaxies in poor
superclusters.  In rich superclusters this scatter is small.

Numerical simulations show that dynamical evolution in high-density regions is
determined by a high overall mean density that speeds up  clustering of
particles (Einasto et al. \cite{e05b} and references therein, Gao, Springel
and White \cite{gsw05}).  In high density regions clustering starts early and
continues until the present.  The haloes that populate high density regions
are themselves also richer, more massive and have larger velocities than the
haloes in low density regions.  In low density filaments that cross voids, as
well as in the outer low density regions of high density systems, the mean
density is low and thus the evolution is slow, and in these regions haloes
themselves are also poor, less massive and have small velocities. These
differences affect the evolution and properties of galaxies in various environments.

In this paper we have used a catalogue of superclusters of galaxies from the 2dF
galaxy redshift survey to study the properties of galaxies in superclusters
and the properties of the richest superclusters.  Our main conclusions are the
following.

\begin{itemize}

  \item{} The density distributions in rich and poor superclusters
  are different. The densities in rich superclusters are higher than 
  in poor superclusters, and rich superclusters contain high density cores which
  are absent in poor superclusters. 

\item{} Rich superclusters contain a higher fraction of early type,
passive, red galaxies than poor superclusters.
  
\item{} The properties of galaxies are correlated with the values 
of the luminosity density field smoothed on a scale of 8 Mpc/h: 
early type,   passive, non-starforming  galaxies have higher 
environmental  densities while late type, active, star/forming
  galaxies have  lower environmental densities.
  This  trend extends to field galaxies and to the lowest densities
  in our sample.
  
\item{}  The fraction of early type, passive galaxies is the highest in rich 
groups/clusters in high density regions of rich superclusters. In these high 
density regions even among isolated galaxies the fraction of star forming galaxies is 
smaller than the fraction of star forming galaxies 
among isolated galaxies in poor superclusters and in 
the field.

\item{} Main galaxies of rich superclusters have larger luminosities (median
  value $-$21.2) than main galaxies of poor superclusters (median value $-$20.9)
  and main galaxies of groups in the field (median value $-$19.7).

\end{itemize}

Our results show that both local (group/cluster)
environments and global (supercluster) environments are  important
in ifluencing galaxy morphologies and their star formation activity. 
This indicates the importance of the role of superclusters, and specially
rich superclusters as a high density environment, which affects the properties of
their member galaxies and groups/clusters of galaxies.

\begin{acknowledgements}
  
  We are pleased to thank the 2dFGRS Team for the publicly available
  data releases. The present study was supported by Estonian Science
  Foundation grants No. 4695, 5347 and 6104, and Estonian Ministry for
  Education and Science support by grant TO 0060058S98. This work has also
  been supported by the University of Valencia through a visiting
  professorship for Enn Saar and by the Spanish MCyT project
  AYA2003-08739-C02-01.  J.E.  thanks Astrophysikalisches Institut Potsdam
  (using DFG-grant 436 EST 17/2/05) for hospitality where part of this study
  was performed. P.H.  was supported by the Finnish Academy of Sciences (grant
  46733).

\end{acknowledgements}

\end{document}